\documentclass[english,aps,prb,preprint,superscriptaddress, group address, amsmath]{revtex4-2}
\usepackage{fancyhdr,graphicx}
\usepackage[table,xcdraw]{xcolor}
\usepackage[normalem]{ulem}
\useunder{\uline}{\ul}{}
\usepackage[T1]{fontenc}
\usepackage{chemformula} 
\usepackage[utf8x]{inputenc}
\usepackage{gensymb}
\usepackage{times}
\usepackage{color}
\usepackage{amsfonts}
\usepackage{amssymb}
\usepackage[colorlinks,bookmarks=false,citecolor=blue,linkcolor=red,urlcolor=blue]{hyperref}
\input pdfcolor.tex
\usepackage{colortbl,amsthm,amsmath,amssymb,txfonts}
\usepackage{graphicx}
\usepackage{epsfig,color}
\usepackage{verbatim}
\usepackage{dcolumn}
\usepackage[normalem]{ulem}
\usepackage{array}
\usepackage{booktabs}
\usepackage{bm}
\usepackage{xcolor}
\usepackage{romannum}

\usepackage{orcidlink}
\usepackage{graphicx} 
\newcounter{suppfigure}

\begin{document}

\title{Layer-Polarization-Driven Metal–Insulator Transition in multi-band Graphene Moir\'e Superlattices}

\author{Harsimran Kaur Mann\orcidlink{0000-0001-7369-550X}$^{1}$, Simrandeep Kaur\orcidlink{0000-0002-1460-0686}$^{2}$, Harsimran Singh\orcidlink{0009-0009-8560-0518}$^1$, Yashashwani Garg$^3$, Amogh Waghmare$^4$, Mohit Kumar Jat\orcidlink{0000-0001-7419-1282}$^1$,  Kenji Watanabe\orcidlink{0000-0003-3701-8119}$^5$, Takashi Taniguchi\orcidlink{0000-0002-1467-3105}$^6$, Manish Jain\orcidlink{0000-0001-9329-6434}$^1$, Aveek Bid\orcidlink{0000-0002-2378-7980}$^{\ast}$ }

\affiliation{Department of Physics, Indian Institute of Science, Bangalore 560012, India\\
$^2$ Fakultät für Physik, Ludwig-Maximilians-Universität, Schellingstraße 4, 80799 München, Germany\\
$^3$ Department of Physical Sciences, Indian Institute of Science Education and Research, Berhampur 760003, India\\
$^4$ Department of Physics, University of
California, Berkeley, California 94720, United States\\
$^5$ Research Center for Electronic and Optical Materials, National Institute for Materials Science, 1-1 Namiki, Tsukuba 305-0044, Japan\\
$^6$ Research Center for Materials Nanoarchitectonics, National Institute for Materials Science, 1-1 Namiki, Tsukuba 305-0044, Japan\\
$^\ast$email: aveek@iisc.ac.in}

\begin{abstract}

Graphene/hBN moir\'e superlattices provide a highly tunable platform for exploring emergent quantum phases in low-dimensional systems. Here, we investigate the moir\'e superlattice formed between hBN and ABA-stacked trilayer graphene (TLG), an inherently multi-band system. We demonstrate that the moir\'e potential is not merely a perturbation but a tool to hybridize the distinct massless and massive electronic sectors of TLG. By applying a perpendicular displacement field to tune layer polarization, we drive a fundamental reconstruction of the electronic band structure. Specifically, increasing the displacement field evolves the system from a multi-band regime to an effectively single-band regime at low energies, accompanied by a metal--insulator transition at the hole-doped secondary Dirac point. This transition originates from a redistribution of carriers across graphene layers that selectively enhances their coupling to the extrinsic moir\'e potential. Quantum capacitance measurements provide direct evidence for the suppression of the density of states at the hole-side secondary Dirac point, consistent with gap opening and the emergence of a displacement-field-tuned band gap. Theoretical calculations reproduce these observations and identify layer-selective coupling to the moir\'e potential as the underlying mechanism. These results demonstrate electrical control of an emergent insulating phase in a low-dimensional moir\'e system, and highlight that layer polarization and layer-selective coupling in multi-band moir\'e heterostructures provide a powerful route for engineering topological and correlated phases through band structure reconstruction and electron interactions.
\end{abstract}

\maketitle

\section{Introduction}

Moir\'e superlattices formed in graphene/hexagonal boron nitride (hBN) heterostructures \cite{PhysRevLett.101.126804, PhysRevB.76.073103, doi:10.1021/nl2005115,xue2011scanning,jat2024higher,yankowitz2012emergence} have emerged as a versatile material platform for realizing a variety of emergent quantum phases. When graphene is closely aligned with hBN or slightly twisted relative to it, the small lattice mismatch gives rise to a long-wavelength moir\'e pattern, which reconstructs the electronic band structure into minibands with reduced bandwidth \cite{yankowitz2012emergence,hunt2013massive,ponomarenko2013cloning,gorbachev2014detecting,woods2014commensurate}. This has enabled the observation of correlated insulating states \cite{chen2020tunable,cao2018correlated,serlin2020intrinsic}, superconductivity \cite{cao2018unconventional,doi:10.1073/pnas.1108174108,balents2020superconductivity}, and various symmetry-broken phases \cite{chen2019signatures,sharpe2019emergent,zhou2024layer,PhysRevLett.133.206504,peng2024abundant}. While such phenomena have been extensively studied in monolayer and bilayer graphene moir\'e systems, these studies have largely focused on regimes where a single set of minibands dominates the low-energy physics, limiting the extent to which electronic structure can be independently controlled. 

Multi-band systems provide an additional degree of tunability arising from the coexistence of bands with different dispersion, symmetry, and layer character. Within this framework, two distinct classes of moir\'e systems emerge: (1) those where the moir\'e potential is \textit{intrinsic}, arising from the mutual twist between identical conducting layers ~\cite{cao2018correlated,serlin2020intrinsic,cao2018unconventional,doi:10.1073/pnas.1108174108,balents2020superconductivity,sharpe2019emergent,PhysRevLett.127.247703,wu2022evidence,doi:10.1126/science.add5574,adak2024tunable}, and (2) those where the potential is \textit{extrinsic}, imposed by a lattice-mismatched substrate such as hBN~\cite{PhysRevLett.101.126804, PhysRevB.76.073103, jat2024higher, yankowitz2012emergence,hunt2013massive, PhysRevB.87.245408,unklapp_mohit, v45r-zcvc, PhysRevB.110.235414}. In the second class of systems, external parameters such as a perpendicular displacement field can selectively modify the relative contribution of these bands at low energies, thereby controlling their hybridization and spectral weight. Crucially, in such systems the moir\'e potential acts in a layer-selective manner, enabling the coupling between electronic states and the moir\'e modulation to be tuned through the spatial distribution of carriers across layers \cite{Moon2014,doi:10.1021/acs.nanolett.3c00253,75gl-jzl6, zibrov2017tunable}. This provides a route to externally control not only the band structure but also the effective strength of the moir\'e potential itself. Such control is particularly relevant in the broader context of emergent quantum phases in low-dimensional systems, where subtle modifications to band structure and interactions can stabilize correlated, topological, or symmetry-broken states. Despite this potential, the interplay between multi-band physics and extrinsic moir\'e superlattices remains comparatively unexplored \cite{mukai2021unconventional}. 

ABA-stacked trilayer graphene (TLG)~\cite{bao2011stacking,lui2011observation,PhysRevB.89.035431,frpm-3fs8, PhysRevB.111.235118, nm8b-5vgm, chanda2025even} is a prototypical graphene-based multi-band system, with a band structure comprising coexisting monolayer-like (MLL) and bilayer-like (BLL) bands (Fig.~\ref{fig:fig1}(a)). These bands are protected by crystal mirror symmetry in the absence of an external electric field. A perpendicular displacement field breaks this symmetry, leading to hybridization between the MLL and BLL bands and allowing continuous tuning between multi-band and effectively single-band regimes~\cite{nm8b-5vgm, PhysRevB.87.115422, PhysRevB.79.125443, PhysRevB.101.245411, PhysRevB.80.195401,PhysRevB.106.205134}. The combination of multi-band structure and layer-selective moir\'e coupling provides a platform in which the coupling to the moir\'e potential can be actively tuned.

In this work, we study an ABA-TLG/hBN moir\'e superlattice and demonstrate that the combined effect of a perpendicular displacement field and the moir\'e potential leads to the emergence of a tunable electronic phase. We find that the displacement field induces strong layer polarization, redistributing carriers across the graphene layers and thereby controlling their coupling to the moir\'e potential, leading to a pronounced asymmetry between electron and hole bands. We observe a displacement-field-driven metal--insulator transition at the hole-side secondary Dirac point, where a finite band gap opens only when carriers are localized in the graphene layer proximal to the moir\'e-forming hBN. In contrast, when carriers reside in layers weakly coupled to the moir\'e potential, the spectrum remains gapless within experimental resolution. Furthermore, we demonstrate that conventional methods for extracting twist angles from secondary Dirac points can fail in multi-band systems due to shifting spectral weights; instead, Brown--Zak oscillations provide a more robust measure of the moir\'e periodicity in these systems. These observations are supported by theoretical calculations that capture the combined effects of interlayer coupling, displacement field, and moir\'e modulation.

Our study establishes multi-band moir\'e systems as a platform where electronic phases can be controlled not only by carrier density and external fields, but also by the spatial distribution of electronic wavefunctions across layers. More broadly, they highlight a general strategy for engineering electronic states in van der Waals heterostructures through layer-selective coupling and band structure control, enabling tunable electronic functionality.
\section{Results}

\begin{figure}[t!]
\includegraphics[width=\columnwidth]{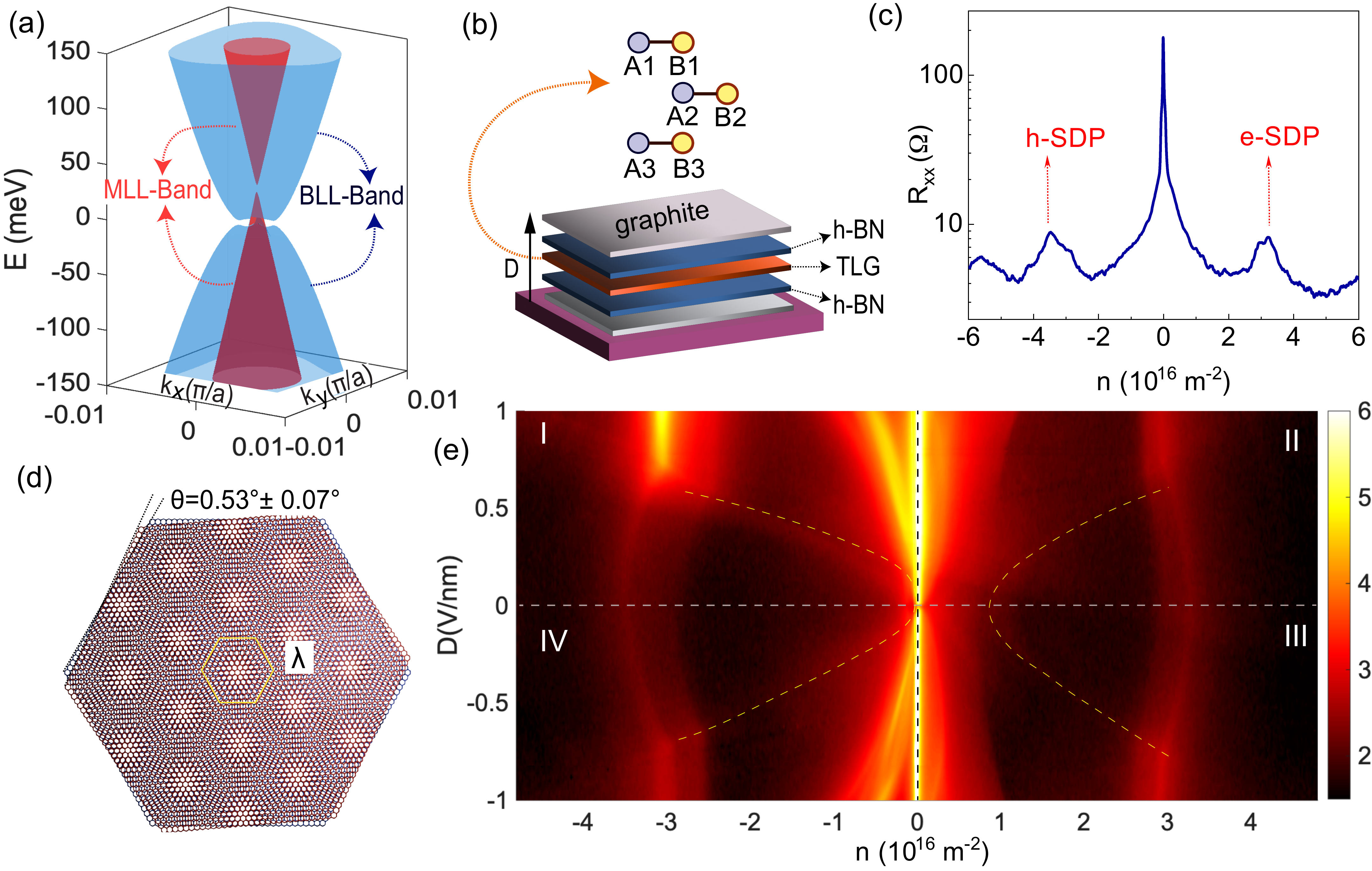}
\caption{\textbf{Device Characterization}: (a) Schematic band structure for ABA TLG at $\Delta=0$~meV, showing the coexistence of MLL and BLL bands. $a=2.46\times10^{-10}~\mathrm{m}$ is the lattice constant. (b) Schematic of the device, defining the direction of positive $D$. ($\mathrm{A1}$ and $\mathrm{B1}$) and ($\mathrm{A3}$ and $\mathrm{B3}$) correspond to the outer graphene layers, while $\mathrm{A2}$ and $\mathrm{B2}$ form the middle layer, shifted relative to the outer layers. The dual-gated geometry enables independent control of $n$ and $D$. (c) Log-scale plot of $R_{xx}$ as a function of $n$. Red vertical arrows at $n=\pm 3.49\times 10^{16}~\mathrm{m^{-2}}$ mark the hole-side secondary Dirac point (h-SDP) and electron-side secondary Dirac point (e-SDP). (d) Schematic of the moir\'e pattern (yellow hexagon) arising from the small angular mismatch between the two hexagonal lattices. (e) Log-scale plot of $R_{xx}$ as a function of $n$ and $D$, measured at $T=20$~mK, illustrating displacement-field tuning of the electronic response. The dotted yellow lines mark the locus of the MLL band in the $n$--$D$ plane. Labels \Romannum{1}--\Romannum{4} denote the different regimes in the phase space.}
\label{fig:fig1}
\end{figure}

High-quality hBN/TLG/hBN dual-graphite-gated devices were fabricated using the dry transfer technique (SM Sec.~1) \cite{Castellanos-Gomez_2014}. The bottom hBN was aligned with the TLG to form a moir\'e superlattice, while the top hBN was intentionally misaligned by a large angle $> 10^\circ$ relative to the TLG (Fig.~\ref{fig:fig1}(b)). This device architecture enables layer-selective coupling to the moir\'e potential while maintaining independent electrostatic control. The dual graphite gates were used to independently tune the carrier density $n$ and the vertical displacement field $D$, providing direct control over both band filling and the layer distribution of electronic states (Fig.~\ref{fig:fig1}(b)). All measurements were performed using standard low-frequency lock-in detection techniques in a dilution refrigerator with a base temperature of $20$~mK. We present the data for \textbf{Device~1}; data for a second device, \textbf{Device~2}, are provided in Sec.~2 of the Supplementary Materials (SM).

Figure~\ref{fig:fig1}(c) shows the longitudinal resistance $R_{xx}$ of \textbf{Device~1} as a function of carrier density. In addition to the peak at the charge neutrality point (CNP), two smaller peaks are observed at $n_s=\pm 3.49\times 10^{16}~\mathrm{m^{-2}}$. These are identified as the hole-side secondary Dirac point (h-SDP) and electron-side secondary Dirac point (e-SDP) (red vertical arrows in Fig.~\ref{fig:fig1}(c)). These additional resistance peaks arise from the formation of a moir\'e superlattice between the TLG and the bottom hBN (Fig.~\ref{fig:fig1}(d)). The twist angle $\theta$ between the TLG and the bottom hBN layer, estimated from the position of the secondary Dirac point, is $0.68^\circ$, whereas that extracted from Brown--Zak oscillations is $0.53^\circ \pm 0.07^\circ$ (SM Sec.~3). This discrepancy is unexpected, as both methods generally yield similar estimates of the twist angle~\cite{unklapp_mohit}. As discussed later, this difference originates from the multi-band nature of the system, where the effective electronic structure and its coupling to the moir\'e potential depend sensitively on carrier distribution across layers.

Figure~\ref{fig:fig1}(e) shows $R_{xx}$ in the $n$--$D$ plane, divided into four regions \Romannum{1}--\Romannum{4}. Two features not typically observed in graphene/hBN moir\'e systems are immediately apparent. First, the magnitude of $R_{xx}$ increases sharply in region \Romannum{1} $(n < 0, D > 0)$ as $D$ exceeds $0.7$~V/nm, indicating a strong enhancement of coupling to the moir\'e potential under displacement-field control. In contrast, in regions \Romannum{2} $(n > 0, D > 0)$, \Romannum{3} $(n > 0, D < 0)$, and \Romannum{4} $(n < 0, D < 0)$, only a weak dependence on $D$ is observed. Second, as $|D|$ increases, the SDP shifts to lower carrier density, reflecting a displacement-field-induced reconstruction of the band structure. Together, these features indicate a pronounced asymmetry in the coupling of carriers to the moir\'e potential, demonstrating that the electronic response of the system can be tuned through controlled redistribution of carriers across graphene layers.

\begin{figure}[t!]
\includegraphics[width=\columnwidth]{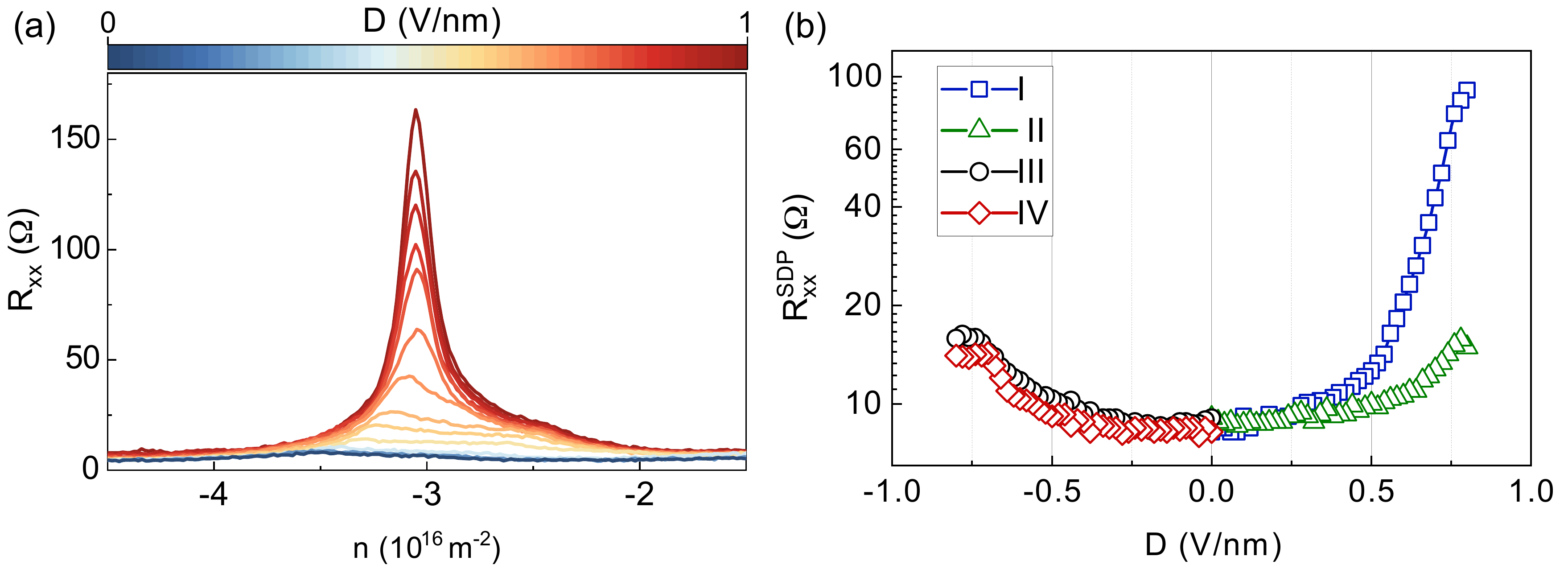}
\caption{\textbf{Layer polarization}: (a) Plot of $R_{xx}$ as a function of $n$ for $0$~V/nm < $D$ < $1$~V/nm and hole-doping (region~\Romannum{1}). (b) Plot of resistance at h-SDP as a function of $D$ in the four regions: Blue open squares -- region~\Romannum{1}, green open triangles -- region~\Romannum{2}, black open circles --region~\Romannum{3}, and red diamonds -- region~\Romannum{4}. The strong asymmetry between positive and negative $D$ highlights the displacement-field-controlled coupling to the moir\'e potential.}
\label{fig:fig2}
\end{figure}

\begin{figure}[t!]
\includegraphics[width=\columnwidth]{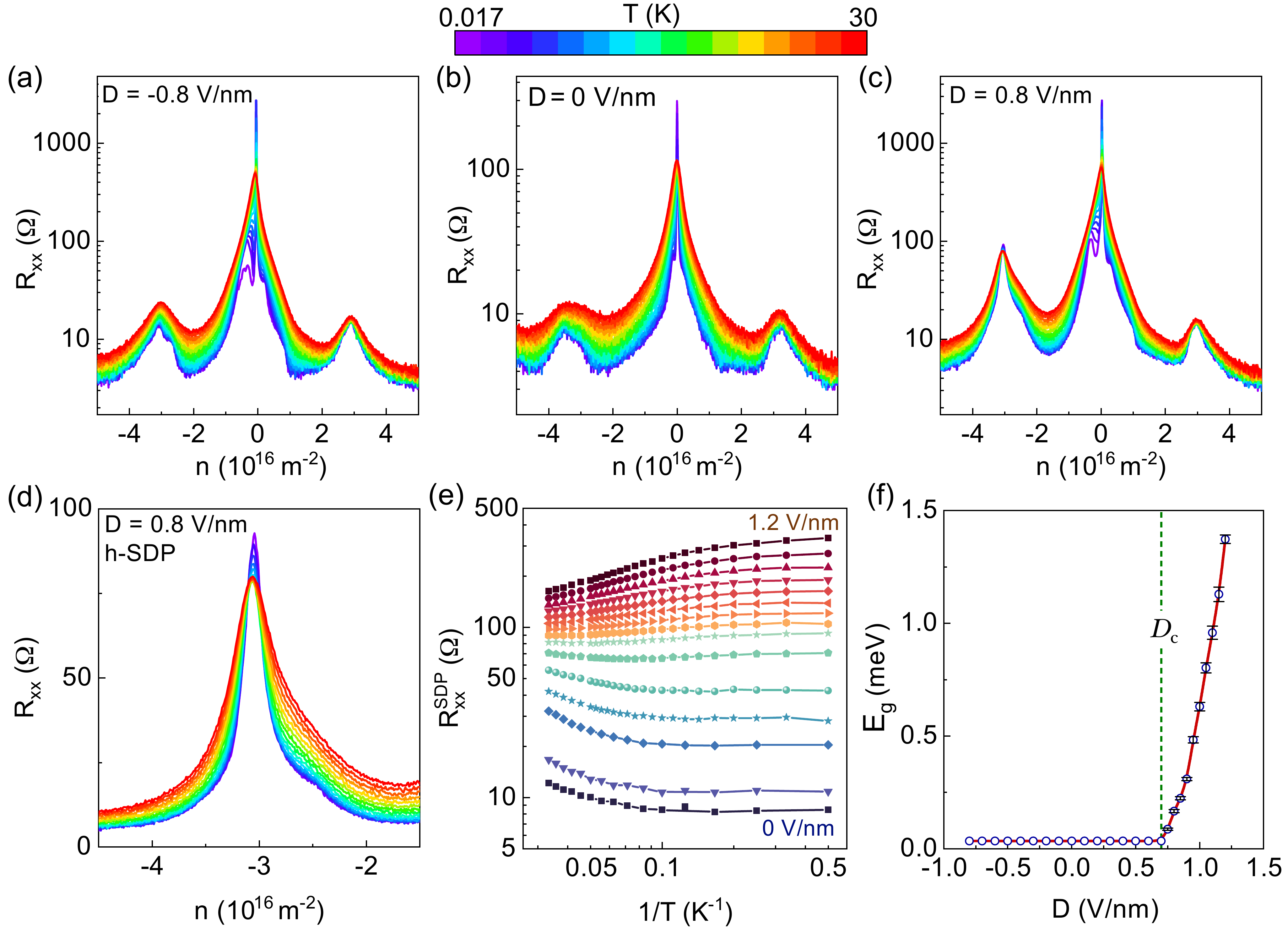}
\caption{\textbf{Displacement-field-tuned band gap}: (a--c) Plots of $R_{xx}$ as a function of $n$ over the temperature range $0.017~\mathrm{K}\leq T \leq 30~\mathrm{K}$ measured at (a) $D=-0.8$~V/nm, (b) $D=0$~V/nm, and (c) $D=0.8$~V/nm, illustrating the evolution of the $T$-dependent electronic response under $D$-field tuning. (d) Zoomed-in plot of $R_{xx}$ as a function of $n$ at the h-SDP for $D=0.8$~V/nm over the range $0.017~\mathrm{K}\leq T \leq 30~\mathrm{K}$. The color scales in (a)--(d) are identical. (e) Plot of $R_{xx}^{SDP}$ at the h-SDP as a function of $1/T$ over the displacement-field range $0~\mathrm{V/nm}<D<1.2~\mathrm{V/nm}$, showing the transition from metallic to insulating behavior. (f) Extracted band gap $E_g$ as a function of $D$, demonstrating a $D$-induced opening of a tunable gap. The red line is a guide to the eye, and the green dotted line marks the critical displacement field $D_c$.}
\label{fig:fig3}
\end{figure}

We now examine these features in detail. Figure~\ref{fig:fig2}(a) shows $R_{xx}$ as a function of $n$ in region \Romannum{1} for representative values of $D$, highlighting the effect of displacement-field control on the electronic response. The resistance at the h-SDP increases from $8~\ohm$ to $164~\ohm$ as $D$ is increased from $0$~V/nm to $1$~V/nm, indicating a strong enhancement of the moir\'e-induced response under positive displacement field. By contrast, $R_{xx}$ is largely insensitive to negative $D$ for h-SDP. (SM Sec.~4), demonstrating a clear asymmetry in the field-controlled behavior. Figure~\ref{fig:fig2}(b) plots the resistance at the SDP, $R_{xx}^{SDP}$, as a function of $D$ across the four regions. The $\sim 20$-fold increase in $R_{xx}^{SDP}$ at the h-SDP for positive $D$ (blue open squares) is not observed in the other three regimes. A pronounced asymmetry between positive (blue open squares) and negative $D$ (red open diamonds) is evident for hole doping (regions~\Romannum{1} and \Romannum{4}), indicating a strong dependence of the moir\'e coupling on the displacement-field-controlled carrier distribution. In contrast, for electron doping (regions~\Romannum{2} and \Romannum{3}), the response remains nearly symmetric in $D$ (green open triangles and black open circles), consistent with a weaker sensitivity of the electronic states to the moir\'e potential (SM Sec.~S9).

Figure~\ref{fig:fig3}(a-c) shows $R_{xx}$ as a function of $n$ over the temperature range $17~\mathrm{mK}<T<30~\mathrm{K}$ for $D=-0.8$~V/nm, $D=0$~V/nm, and $D=0.8$~V/nm, respectively, illustrating the effect of displacement-field tuning on the temperature-dependent electronic response. For $D=-0.8$~V/nm and $D=0$~V/nm, $R_{xx}$ increases with temperature ($dR_{xx}/dT>0$) for both the h-SDP and e-SDP, indicating that the Fermi level does not lie within a significant band gap requiring thermal activation of carriers. In contrast, for $D=0.8$~V/nm, while $dR_{xx}/dT>0$ for $n>0$, $dR_{xx}/dT<0$ for $n<0$, indicating insulating behavior for hole doping. This is further evident from Fig.~\ref{fig:fig3}(d), which shows a zoomed-in view in the regime $D=0.8$~V/nm, $n<0$ (region~\Romannum{1}). Thus, not only the magnitude of $R_{xx}(D)$, but also its temperature dependence is qualitatively distinct in region~\Romannum{1} compared to the other three regions, demonstrating electrical control over the insulating state through the displacement field. This behavior signifies the emergence of a displacement-field-tunable insulating phase in the moir\'e system.

Figure~\ref{fig:fig3}(e) shows $R_{xx}^{SDP}$ at the h-SDP as a function of $1/T$ for different values of $D$. At high temperatures (small $1/T$), the temperature dependence changes sign at a critical displacement field $D_c \sim 0.7$~V/nm, indicating a transition from metallic $(D < D_c)$ to insulating $(D > D_c)$ behavior under displacement-field tuning. The band gaps $E_g$ in the insulating regime were extracted using an Arrhenius fit to the $R_{xx}^{SDP}$–$T$ data, $R_{xx}^{SDP}=R_0 \exp\left(E_g/2k_B T\right)$~\cite{nm8b-5vgm} for $T \geq 4~\mathrm{K}$ (transport is dominated by variable range hopping below this temperature), and are plotted in Fig.~\ref{fig:fig3}(f). The extracted $E_g$ at the h-SDP remains below experimental resolution for $D \leq 0.7$~V/nm, after which it increases, reaching $E_g \sim 1.4$~meV at $D=1.2$~V/nm, demonstrating a displacement-field-induced opening of a tunable band gap. 

\begin{figure}[t!]
\includegraphics[width=\columnwidth]{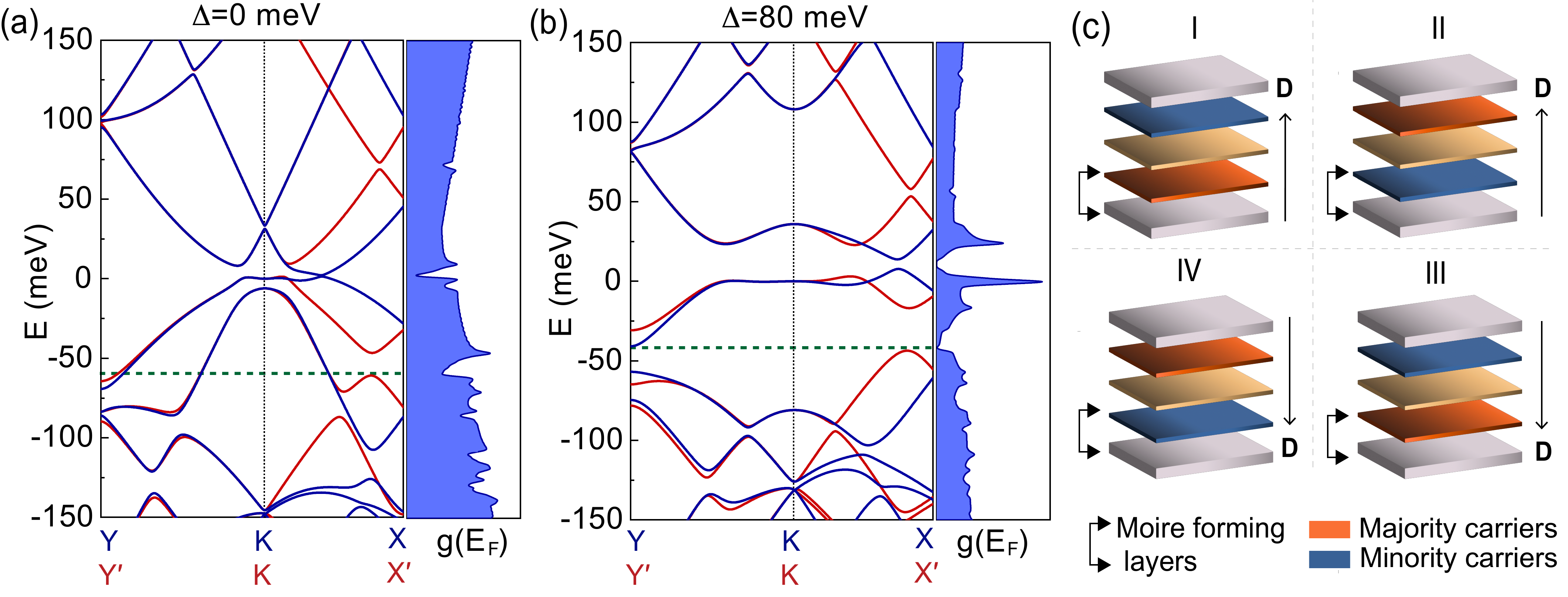}
\caption{\textbf{Calculated band structure}: Calculated band structure of the TLG/hBN moir\'e heterostructure for $\theta = 0.55^\circ$ for (a) $\Delta=0$~meV and (b) $\Delta=80$~meV, shown within a single valley. The blue (red) curves correspond to the high-symmetry path $\mathrm{Y} \rightarrow \mathrm{K} \rightarrow \mathrm{X}$ ($\mathrm{Y'} \rightarrow \mathrm{K} \rightarrow \mathrm{X'}$). The blue shaded regions represent the density of states $g(E_F)$ as a function of energy $E$ in arbitrary units, highlighting the modification of the electronic structure under displacement-field tuning. (c) Schematic of the TLG/hBN moir\'e device illustrating the $D$-controlled distribution of majority and minority carriers across the graphene layers. Grey planes represent hBN, while blue, yellow, and orange planes represent the three graphene layers of TLG. The orange layer denotes the layer hosting the majority carriers, while the blue layer represents the minority carrier contribution.}
\label{fig:fig4}
\end{figure}

To summarize our primary observations, we identify four key experimental features. First, the longitudinal resistance exhibits a pronounced asymmetry with respect to displacement field and carrier polarity, with a sharp enhancement at the hole-side secondary Dirac point only for positive displacement field (region I), while remaining weakly dependent on $D$ in the other regimes. Second, the position of the secondary Dirac point shifts systematically to lower carrier density with increasing $|D|$, indicating a reconstruction of the underlying band structure. Third, temperature-dependent transport reveals a clear metal--insulator transition at the h-SDP above a critical displacement field $D_c \sim 0.7~\mathrm{V/nm}$, accompanied by the opening of a finite band gap, whereas no such transition is observed on the electron side. Finally, we observe a discrepancy between the twist angle extracted from the position of the resistance peaks and that obtained from Brown--Zak oscillations, reflecting the breakdown of a single-band description in this multi-band system. Taken together, these observations establish the displacement field as a control parameter for accessing distinct emergent electronic phases in the moir\'e superlattice.

To understand these observations, we compute the band structure of TLG under the simultaneous influence of $D$ and the moir\'e superlattice potential. 
For this, we consider a continuum model for ABA-stacked trilayer graphene \cite{Serbyn2013,Taychatanapat2011,Dresselhaus2002, PhysRevB.81.115315, PhysRev.109.272}, valid for small twist angles, incorporating all relevant intra- and interlayer hopping processes together with a layer--asymmetric potential induced by a perpendicular displacement field $\Delta$ \cite{PhysRevB.79.125443} (SM Sec.~S7). 
The presence of hBN moir\'e is accounted for by integrating out the hBN degrees of freedom \cite{Moon2014, PhysRevB.86.115415, PhysRevB.87.245408, PhysRevB.87.205404}, resulting in an effective moir\'e potential acting on one of the graphene layers. The parameters of this potential, including the uniform component $V_0$ and the first-harmonic amplitude $V_1$ and phase $\psi$, are obtained from an electric-field-dependent formulation in which $\Delta$ modifies the effective boron and nitrogen onsite energies \cite{PhysRevB.86.115415}. This formalism provides a unified description of the combined effects of interlayer coupling, external electric field, and substrate-induced moir\'e modulation (see SM Sec.~S7).

Our calculations show that at zero perpendicular displacement field ($\Delta = 0$~meV), the effect of the moir\'e potential is strongly asymmetric in energy, being significantly more pronounced for holes than for electrons (Fig.~\ref{fig:fig4}(a)). For hole doped part of the spectrum, the moir\'e potential substantially reconstructs the density of states $g(E_F)$, leading to the opening of a gap at the h-SDP whose magnitude increases with $|D|$ (see SM Fig.~S7). In contrast, the electron-side density of states remains largely unaffected by the moir\'e potential (Fig.~\ref{fig:fig4}(b)). These results, together with the fact that the moir\'e potential in our device primarily acts on the graphene layer adjacent to the bottom hBN, provide a natural explanation for the observations in Fig.~\ref{fig:fig2}. 

Fig.~\ref{fig:fig4}(c) shows a schematic illustrating the distribution of charge carriers across the layers in the different regimes. In regime~\Romannum{1} ($D>0$, $n<0$), the displacement field $D$ drives holes (the majority carriers) towards the bottom graphene layer, consistent with the calculated layer-resolved weight of the electronic states (see SM Fig.~S9). Since this layer is proximate to the aligned hBN, the holes experience a strong moir\'e potential, resulting in the opening of a gap at the h-SDP. As $D$ increases, the layer polarization becomes stronger, increasing the weight of the hole wavefunction on the bottom layer and leading to a corresponding increase in $R^{SDP}_{xx}$ and $E_g$. 
By contrast, in regime~\Romannum{4} ($D<0$, $n<0$), the displacement field drives holes towards the top graphene layer, away from the moir\'e potential; consequently, the spectrum remains gapless and $R_{xx}$ shows only a weak dependence on $D$. In regimes~\Romannum{2} and \Romannum{3} ($n>0$ and $\pm D$), electrons are the majority carriers. Since the positive-energy spectrum is only weakly perturbed by the moir\'e potential, the resistance remains nearly symmetric in $\pm D$, varies weakly with $|D|$ (see SM Fig.~S7), and no measurable bandgap opens at the e-SDP.

\begin{figure}[t!]
\includegraphics[width=\columnwidth]{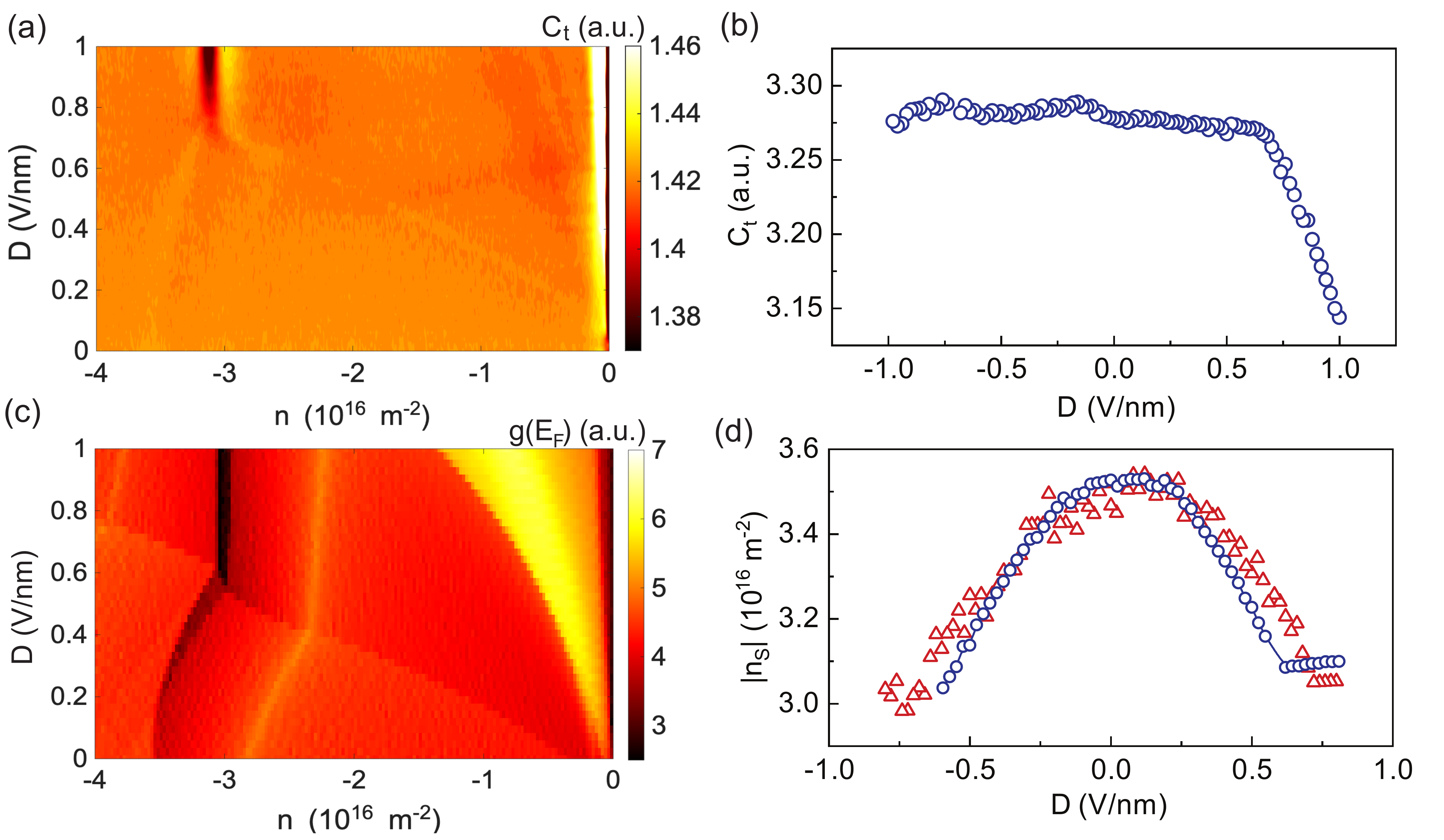}
\caption{\textbf{Quantum capacitance as a direct probe of displacement-field-tunable density of states}: (a) Measured total capacitance $C_t$ as a function of $n$ and $D$, reflecting the evolution of the density of states under displacement-field tuning. (b) Measured total capacitance $C_t$ as a function of $D$ at h-SDP. (c) Calculated $g(E_F)$ as a function of $n$ and $D$, capturing the displacement-field-induced modification of the electronic structure. (d) Position of the secondary Dirac point, $n_s$, as a function of $D$, extracted from experimental data (open red triangles) and calculated data (blue open circles) (see Sec.~S6 of the SM for details).}
\label{fig:fig5}
\end{figure}

To verify the origin of the band gap in region~\Romannum{1}, we measured the quantum capacitance, which directly probes the density of states at the Fermi level, $C_Q = e^2 g(E_F)$ (SM Sec.~5). The experimentally measured quantity is the total capacitance, $C_t = \left(1/C_Q  + 1/C_g\right)^{-1}$,  where $C_g$ is the geometrical gate capacitance. 
Figure~\ref{fig:fig5}(a) shows a contour plot of the measured $C_t$ as a function of $D$ and $n$ (data over the full $n$--$D$ range are presented in SM Sec .~5). Since $C_t$ includes both geometric and quantum contributions, it is sensitive to variations in $g(E_F)$. Figure~\ref{fig:fig5}(b) shows a line cut of $C_t$ as a function of $D$ at h-SDP. We find that $C_t$ remains nearly unchanged over the range $-1~\mathrm{V/nm} < D < 0.7~\mathrm{V/nm}$, before decreasing sharply for $D \geq 0.7$~V/nm, signaling a reduction in the density of states (Fig.~S6 of SM).  This pronounced asymmetry between positive and negative $D$ indicates that a band gap opens only when holes are localized in the graphene layer adjacent to the hBN that forms the moir\'e potential. 
The capacitance measurements thus provide direct evidence for a displacement-field-tunable insulating phase.
Consistent with transport measurements and our calculations, this behavior arises from a displacement-field-induced redistribution of carriers across layers, which selectively enhances their coupling to the moir\'e potential. Figure~\ref{fig:fig5}(c) shows the calculated $g(E_F)$ as a function of $D$ and $n$, in good agreement with the experimental data.

Additionally, as seen in both the experimental and theoretical data (Fig.~\ref{fig:fig5}(a,c)), the position of the h-SDP shifts to lower carrier density as $D$ is increased from $0$ to $0.7$~V/nm, before saturating at $n = 3.05 \times 10^{16}~\mathrm{m^{-2}}$ for $D \geq 0.7$~V/nm. The quantitative agreement between experiment and theory is shown in Fig.~\ref{fig:fig5}(d), which plots $n_s$ as a function of $D$ for both calculated (blue open circles) and experimental (red open triangles) data. 

This evolution of the h-SDP with $D$ can be understood from the displacement-field-induced shift of the MLL band to higher energies. As a result of this migration, the contribution of the MLL band to the density of states near the SDP is progressively reduced, leading to the observed shift in $n_s$. Once the MLL band is no longer relevant (for $D > D_c$) and only the BLL band contributes, the system enters an effectively single-band regime in the density range of interest, and the position of the SDP becomes nearly constant.

This $D$-induced transition from a multi-band to an effectively single-band regime also explains the discrepancy between the twist angle $\theta$ obtained from Brown--Zak oscillations and that extracted from the position of $n_s$ at $D = 0$~V/nm. Brown--Zak oscillations determine $\theta$ from the flux through a moir\'e unit cell and are therefore independent of $D$. In contrast, extracting $\theta$ from the position of resistance maxima at the SDP assumes that the corresponding carrier density is set by a single-band filling condition, which does not hold in a multi-band system such as ABA-TLG. As a result, the two methods yield different values of $\theta$ in the multi-band regime. Notably, the value of $\theta$ obtained from Brown--Zak oscillations agrees well with that extracted from the SDP position for $D \geq 0.7$~V/nm, where a single BLL band dominates transport. These results highlight that Brown--Zak oscillations provide a more reliable measure of the twist angle in multi-band moir\'e systems.

\section{Conclusion}

Our measurements establish that in an ABA-TLG moir\'e heterostructure, the moir\'e potential couples strongly and asymmetrically to holes and electrons, leading to the emergence of a displacement-field-controlled electronic phase. We show that a perpendicular displacement field drives a transition from a multi-band metallic state to a single-band insulating state by inducing pronounced layer polarization. At high $D$, holes are localized in the graphene layer adjacent to the hBN that forms the moir\'e potential, leading to an electric field-controlled metal--insulator transition at the h-SDP for $D \geq D_C$, accompanied by the opening of a finite band gap. Theoretical calculations quantitatively reproduce these observations and identify layer-selective coupling to the moir\'e potential as the underlying mechanism. Quantum capacitance measurements provide independent confirmation by directly probing the density of states, revealing a clear suppression at the h-SDP for $D \geq D_C$ consistent with gap opening. We further establish that Brown--Zak oscillations provide a robust and reliable measure of the twist angle in multi-band materials such as ABA-TLG, where conventional extraction from secondary Dirac point positions breaks down.

Layer polarization effects have been explored previously in moir\'e systems, particularly in twisted graphene structures, where they influence correlated phases by redistributing carriers between layers within an intrinsically formed moir\'e potential. In contrast, the mechanism demonstrated here is fundamentally distinct: the moir\'e potential is extrinsic and acts predominantly on a single graphene layer, enabling displacement-field control over the coupling between electronic states and the moir\'e modulation itself. As a result, layer polarization does not merely redistribute carriers within a fixed moir\'e landscape, but actively tunes their exposure to the moir\'e potential, driving a transition between weakly and strongly coupled regimes. This establishes a distinct mode of control in which the effective strength of the moir\'e potential can be modulated through an external electric field.

Our results establish ABA-TLG/hBN moir\'e as a tunable platform in which the interplay of moir\'e potential and displacement field enables controlled manipulation of electronic states and phase transitions. More broadly, they identify layer-selective coupling as a general route for engineering electronic phases in multi-band van der Waals heterostructures, and establish it as a scalable design principle for achieving electrically tunable functionality in low-dimensional quantum materials.

\textbf{Acknowledgments} \par 
A.B. acknowledges funding from the Department of Science and Technology, Govt of India (SP/ANRF-24-0117). K.W. and T.T. acknowledge support from the JSPS KAKENHI (Grant Numbers 21H05233 and 23H02052) and World Premier International Research Center Initiative (WPI), MEXT, Japan. M.J. acknowledges funding from the Nano mission of the Department of Science and Technology, India (DST/NM/TUE/QM-10/2019). M.K.J. acknowledges funding from the Prime Minister's research fellowship (PMRF).

\clearpage

\section*{Supplementary information}

\renewcommand{\theequation}{S\arabic{equation}}
\renewcommand{\thesection}{S\arabic{section}}

\renewcommand{\thetable}{{S\arabic{table}}}

\setcounter{equation}{0}
\setcounter{section}{0}
\setcounter{table}{0}

\section{Device fabrication}

Bernal-stacked trilayer graphene (TLG) and hexagonal boron nitride (hBN) flakes were mechanically exfoliated onto $\ch{Si}/\ch{SiO2}$ substrates. The hBN flakes had a thickness of 25--30~nm. The TLG flakes were initially identified from optical contrast under a microscope and subsequently confirmed using Raman spectroscopy. The flakes were assembled into a stack using a polycarbonate (PC) film in the sequential order of graphite/hBN/TLG/hBN/graphite. The TLG was aligned to the bottom hBN~\cite{unklapp_mohit}. The stack was then transferred onto a $\ch{Si}/\ch{SiO2}$ substrate along with the PC film, followed by removal of the PC residue in chloroform. The heterostructure was subsequently annealed in vacuum at $300^\circ$~C for 4~hours. Electron-beam lithography was used to define the contacts, followed by etching with a mixture of $\ch{CHF_3}$ (40~sccm) and $\ch{O_2}$ (10~sccm). Metallization was carried out using $\ch{Cr/Pd/Au}$ (3~nm/12~nm/55~nm) to form one-dimensional electrical contacts to the TLG~\cite{kaur_universality_2024}.

The dual graphite gates were used to tune carrier density $n$ and displacement field $D$ simultaneously using $n=(C_{bg}V_{bg}+C_{tg}V_{tg})/e+n_0$ and $D=(C_{bg}V_{bg}-C_{tg}V_{tg})/2\epsilon_0+D_0$. Here $C_{bg}$($C_{tg}$) is the back-gate (top-gate) capacitance and $V_{bg}$($V_{tg}$) is the back-gate (top-gate) voltage. $n_0$ and $D_0$ are the residual number density and displacement field due to impurities in graphene.  

 The field-effect mobility $\mu$ of the devices was extracted by fitting $R_{xx}$ measured at $D=0~\mathrm{V/nm}$ and $B=0$~T to the expression $R = R_c + {L}/({W e \mu \sqrt{n^2 + n_0^2}})$. Here, $R_c$, $L$, and $W$ are the contact resistance, length, and width of the device, respectively. From the fit, we estimate $\mu \sim 62~\mathrm{m^2 V^{-1} s^{-1}}$ and $n_0 \sim 7.81 \times 10^{13}~\mathrm{m^{-2}}$ for \textbf{Device~1}. The corresponding numbers for \textbf{Device~2} are $\mu \sim 90~\mathrm{m^2 V^{-1} s^{-1}}$ and $n_0 \sim 1.2\times10^{14}~\mathrm{m^{-2}}$.

\section{Data for Device 2 \label{sec:device2}}

\begin{figure}[h]
    \centering
    \includegraphics[width=1\columnwidth]{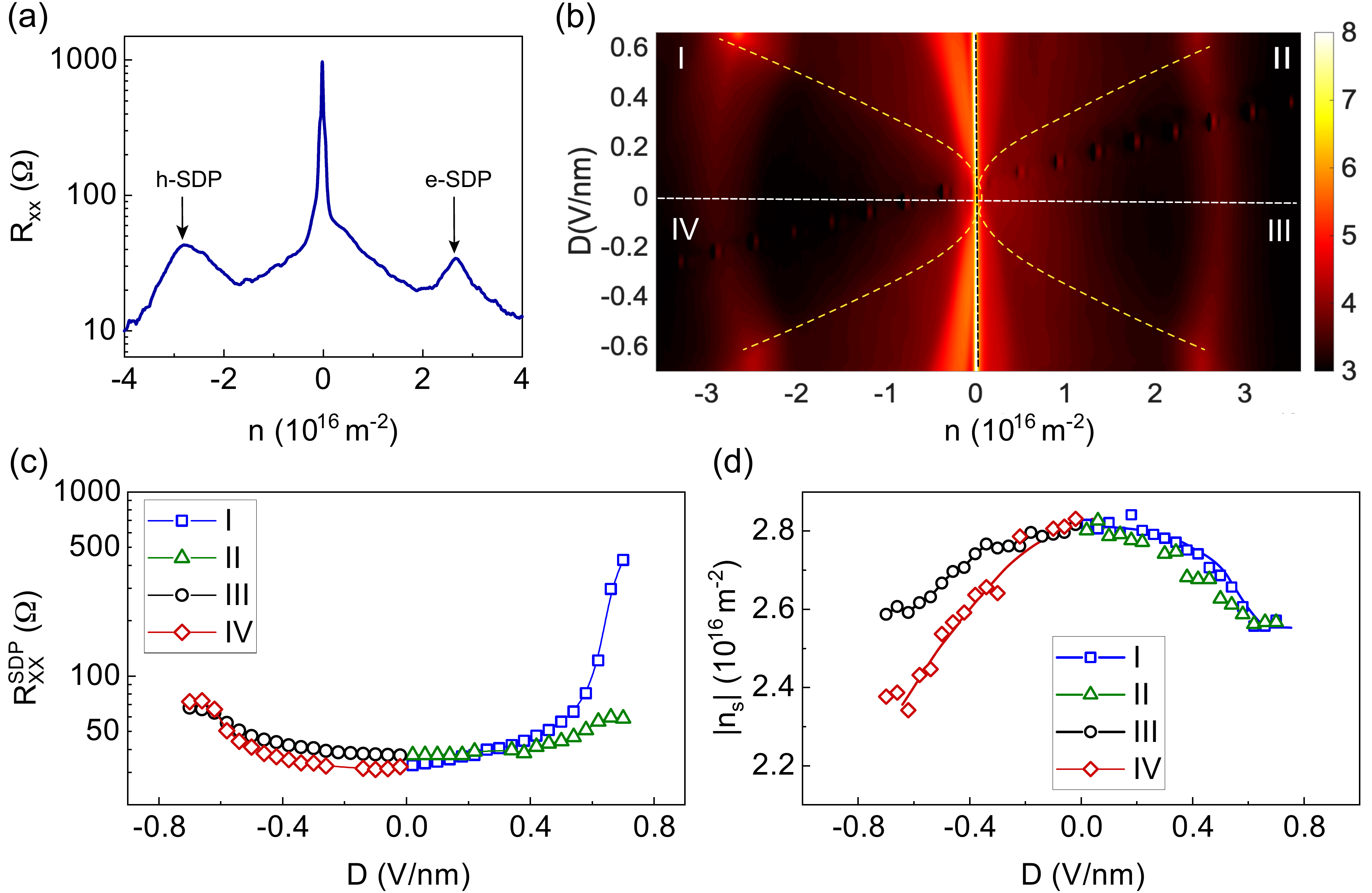}
    \caption{(a) Log-scale plot of $R_{xx}$ as a function of $n$. (b) Plot of $R_{xx}$ on a logarithmic scale as a function of $n$ and $D$. Yellow dotted lines represent the monolayer-like (MLL) band. (c) Plot of $R_{xx}$ at SDP as a function of $D$ for regions \Romannum{1}-\Romannum{4} . (d) Plot of $|n_s|$ as a function of $D$ for regions \Romannum{1}-\Romannum{4} .   Region~\Romannum{1}--blue open squares, region~\Romannum{2}--green open triangles, region~\Romannum{3}--black open circles, and region~\Romannum{4}--red open diamonds. The lines in (c) and (d) are guides to the eye. }
    \label{figS1}
\end{figure}

Fig.~\ref{figS1}(a) shows $R_{xx}$ as a function of $n$ for \textbf{Device~2}. The secondary Dirac points appear at $n_s= \pm 2.8\times10^{16}~\mathrm{m^{-2}}$. The measured twist angle between hBN and TLG is $\theta = 0.28^\circ$. Fig.~\ref{figS1}(b) shows a contour plot of $R_{xx}$ as a function of $n$ and $D$ for \textbf{Device~2}. Similar to \textbf{Device~1}  (data in the main manuscript), we observe two prominent features: 
\begin{enumerate}
    \item The magnitude of $R_{xx}$ increases sharply in region \Romannum{1} as $D$ is increased above $0.6$~V/nm, whereas for regions \Romannum{2}, \Romannum{3}, and \Romannum{4}, only a small increase in resistance is observed for finite, large $D$. 
    \item As $|D|$ increases, the SDP shifts to lower carrier density.
\end{enumerate}

Fig.~\ref{figS1}(c) shows a plot of $R^{SDP}_{xx}$ for regions \Romannum{1}-\Romannum{4} as a function of $D$. An asymmetry is observed between $D>0$ and $D<0$ at the h-SDP, whereas for e-SDP it remains largely symmetric. 
On the hole side, $R^{SDP}_{xx}$ increases sharply from $32~\ohm$ to $425~\ohm$ as $D$ is increased from $0$ to $0.6~\mathrm{V/nm}$ (blue open squares). In contrast, when $D$ is swept in the negative direction from $0$ to $-0.6~\mathrm{V/nm}$, the resistance shows only a small increase from $32~\ohm$ to $64~\ohm$ (red open diamonds). This behavior is consistent with \textbf{Device~1} (data presented in the main manuscript). 
 
Fig.~\ref{figS1}(d) shows a plot of the magnitude of $n_s$ as a function of $D$ for regions \Romannum{1}-\Romannum{4}. SDP shifts to lower carrier density as $D$ is increased. These results again follow the observations made for \textbf{Device~1} (data presented in the main manuscript).

\section{Estimating the twist angle}

\begin{figure}[h]
    \centering
\includegraphics[width=1\columnwidth]{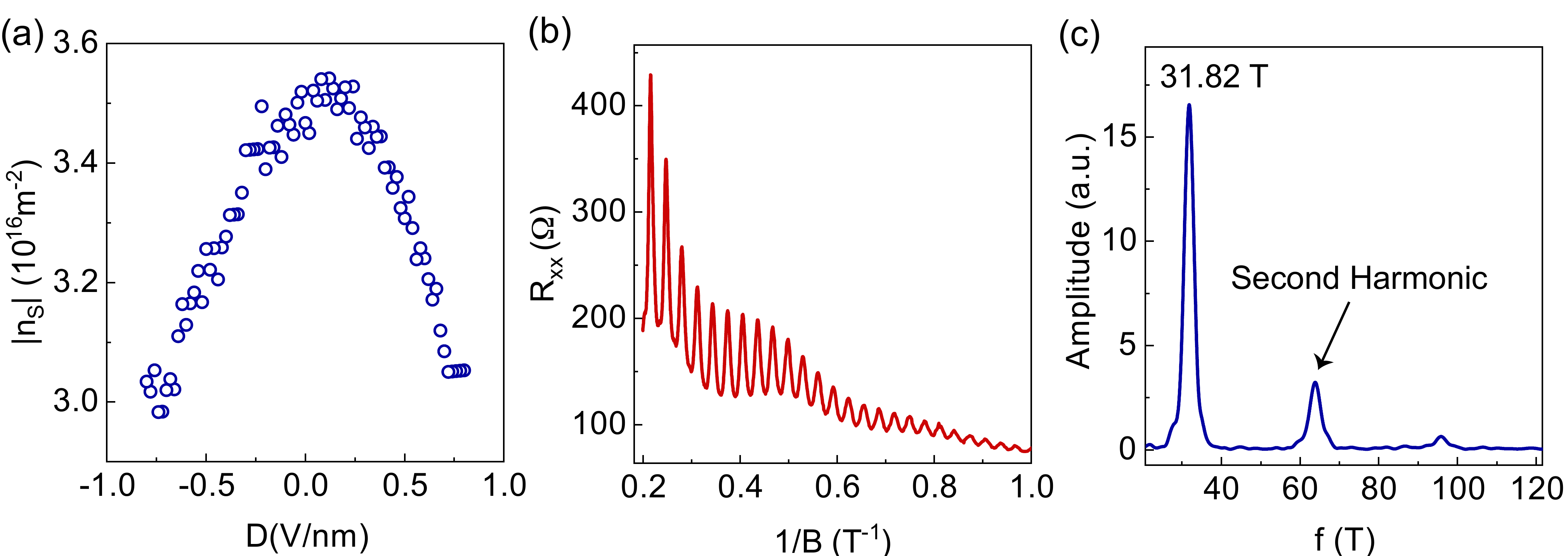}
    \caption{(a) $|n_s|$ as a function of $D$. (b) $R_{xx}$ as a function of $1/B$ at $n=2.8\times 10^{16}~\mathrm{m^{-2}}$ and $T=30~\mathrm{K}$. (c) Fast Fourier transform of the data in panel (b). The primary oscillation frequency $B_f = 31.82$~T and its second harmonic are marked. }
    \label{figS2}
\end{figure}

The twist angle of the device can be estimated using two methods: (1) from the carrier density at which the secondary Dirac point appears, and (2) from Brown--Zak oscillations.

\textbf{Method~1:} Fig.~\ref{figS2}(a) shows the magnitude of the carrier density $n_s$ at the h-SDP as a function of $D$. As $|D|$ increases from $0~\mathrm{V/nm}$, $n_s$ decreases continuously. The corresponding moir\'e wavelength $\lambda$ is estimated using the relation~\cite{doi:10.1126/sciadv.abd3655}:
\begin{equation}
    \lambda^2 = \frac{8}{\sqrt{3}n_s}
\end{equation}
The calculated values of $\lambda$ for $D_1 = 0$~V/nm, $D_2 = 0.4$~V/nm, $D_3 = 0.6$~V/nm, and $D_4 = 0.8$~V/nm are $\lambda_1 = 11.49$~nm, $\lambda_2 = 11.66$~nm, $\lambda_3 = 11.92$~nm, and $\lambda_4 = 12.31$~nm, respectively.

\textbf{Method~2:} Fig.~\ref{figS2}(b) shows the magneto resistance oscillations at $30~\mathrm{K}$. At elevated temperatures, the Landau levels are smeared out, and only the magneto transport oscillations arising from recurring Bloch states in the superlattice persist. These are known as Brown--Zak oscillations. The fast Fourier transform of the data yields a single frequency $B_f = 31.82\pm 3~\mathrm{T}$ as shown in Fig.~\ref{figS2}(c). This frequency is related to the real-space area $S$ of the moir\'e unit cell by $B_f = \phi_0/S$, where $\phi_0 = h/e$ is the magnetic flux quantum~\cite{PhysRevB.14.2239,doi:10.1126/science.aal3357}. Using
\begin{equation}
    \lambda_{BZ} = \sqrt{\frac{2S}{\sqrt{3}}},
\end{equation}
the calculated value of the moir\'e wavelength is $\lambda_{BZ} = 12.25\pm0.58$~nm. This value is independent of $D$ and depends only on the geometry of the real-space lattice.

Note that $\lambda_{BZ} \simeq \lambda(D = 0.8~\mathrm{V/nm})$. This implies that the moir\'e wavelength extracted from resistance maxima in multiband systems may be unreliable and yields the correct value only in the limit where a single band crosses the Fermi energy.

We estimate the twist angle magnitude between TLG and hBN using the relation~\cite{doi:10.1021/acs.nanolett.8b05061}:
\begin{equation}
    \lambda = \frac{(1+\epsilon)a}{[\epsilon^2 + 2(1+\epsilon)(1 - \cos(\theta))]^{1/2}}
\end{equation}
Here $a = 0.246$~nm is the lattice constant of graphene, $\epsilon = 0.0182$ is the lattice mismatch between hBN and graphene, and $\theta$ is the relative twist angle between hBN and TLG. Using $\lambda = 12.25\pm 0.58$~nm, we obtain $\theta = 0.53^\circ\pm 0.07^\circ$.

\section{$D$-dependent measurement}

\begin{figure}[h]
    \centering
    \includegraphics[width=1\columnwidth]{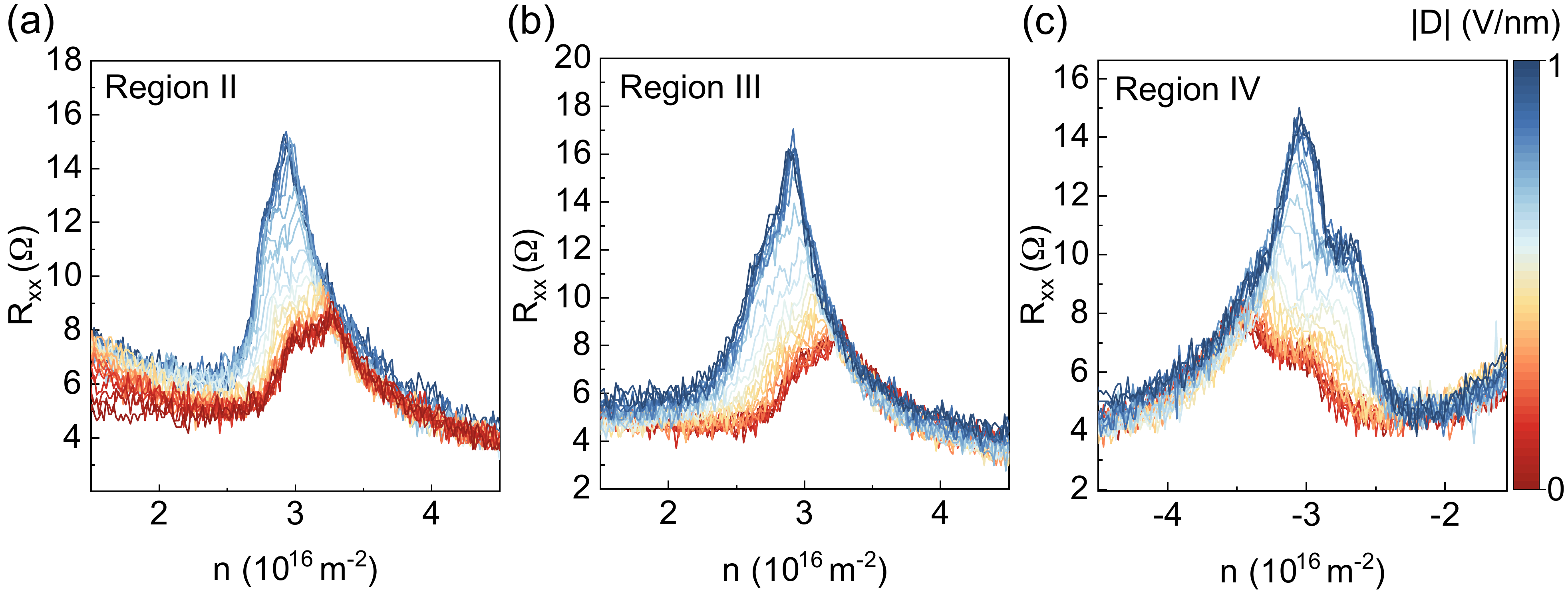}
    \caption{$R_{xx}$ as a function of $n$ for $0 < |D| < 1~\mathrm{V/nm}$ in (a) region \Romannum{2}, (b) region \Romannum{3}, and (c) region \Romannum{4}.}
    \label{figS3}
\end{figure}

Fig.~\ref{figS3}(a)--(c) show $R_{xx}$ as a function of $n$ for regions \Romannum{2}--\Romannum{4} of Fig.~1(e) in the main manuscript. We find that, with increasing $|D|$, the increase in $R_{xx}$ at the SDP in all three regions is much smaller than that observed in region \Romannum{1}. A quantitative comparison of the change in $R_{xx}^{SDP}$ across different regions is summarized in Table~\ref{table:comparison}.

\begin{table}[h]
\centering
\begin{tabular}{cccc}
\hline
Region & Carrier type & $\Delta R_{xx}^{SDP}$ ($\Omega$) & Relative change \\
\hline
\Romannum{1} & Holes ($D>0$) & $\sim 150$ & $\sim 20\times$ \\
\Romannum{2} & Electrons ($D>0$) & $\sim 10$ & $\sim 2\times$ \\
\Romannum{3} & Electrons ($D<0$) & $\sim 5$--$10$ & $\sim 2\times$ \\
\Romannum{4} & Holes ($D<0$) & $\sim 5$--$10$ & $\sim 2\times$ \\
\hline
\end{tabular}
\caption{Quantitative comparison of the change in $R_{xx}$ at the secondary Dirac point (SDP) across different regions as a function of displacement field $D$. \label{table:comparison}}
\end{table}

\section{Quantum capacitance measurements}

\begin{figure}[b]
    \centering
    \includegraphics[width=1\columnwidth]{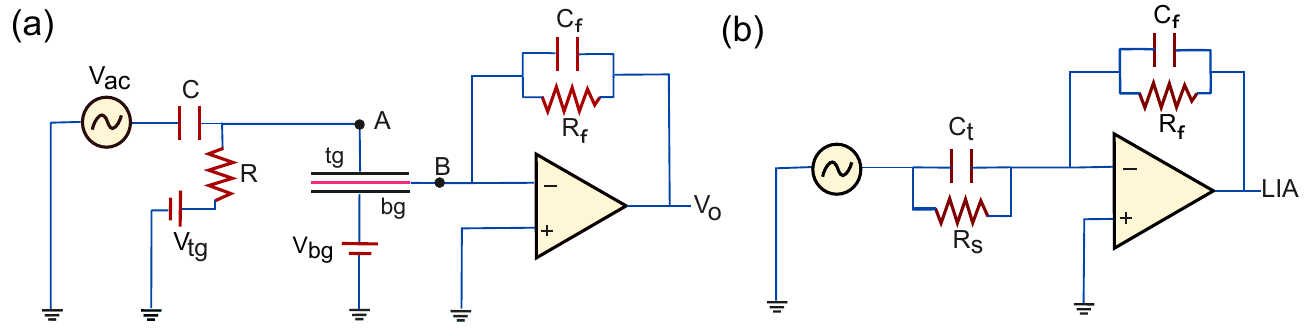}
    \caption{(a) Device schematic of the measurement circuit. $\mathrm{V_{bg}}$ and $\mathrm{V_{tg}}$ are the DC voltages applied to the back gate (bg) and top gate (tg), respectively. The feedback loop consists of $\mathrm{C_f} = 3.9 \times 10^{-12}~\mathrm{F}$ and $\mathrm{R_f} = 5.6 \times 10^{3}~\Omega$ connected in parallel. The TLG channel receives the input signal from the mixer circuit ($\mathrm{C} = 22 \times 10^{-9}~\mathrm{F}$, $\mathrm{R} = 100 \times 10^{3}~\Omega$) at node A. (b) An effective circuit where the device is modeled as a leaky capacitor with total capacitance $\mathrm{C_t}$ and resistance $\mathrm{R_s}$. The output is measured using a lock-in amplifier (LIA). 
    \label{figS4}}
\end{figure}

The schematic of the circuit used for capacitance measurements is shown in Fig.~\ref{figS4}(a). An operational amplifier (op-amp) is used as a transimpedance amplifier to convert the current signal into an output voltage $\mathrm{V_o}$, whose in-phase and out-of-phase components are measured using lock-in amplifiers from which the capacitance is extracted.

A passive mixer circuit is used to superimpose an AC signal $\mathrm{V_{ac}} = 20$~mV at $50$~kHz on the DC bias $\mathrm{V_{tg}}$ applied to the top gate. This signal is passed through the device, which is held at virtual ground through the inverting input of the op-amp, and the output voltage is measured. A negative feedback loop consisting of a parallel combination of a capacitor $\mathrm{C_f} = 3.9 \times 10^{-12}~\mathrm{F}$ and a resistor $\mathrm{R_f} = 5.6 \times 10^{3}~\Omega$ is employed. The device is modeled as a leaky capacitor, and the corresponding effective circuit is shown in Fig.~\ref{figS4}(b). From the measured output voltage $\mathrm{V_o}$, the capacitance is extracted using
\begin{equation}
    \frac{V_{ox} + j V_{oy}}{V_{Ax} + j V_{Ay}} = -\frac{R_f + {1}/{(j \omega C_f)}}{R_s +{1}/{(j \omega C_t)}},
\end{equation}
where $C_t$ is the total capacitance and $R_s$ is the resistance associated with the device. $V_{ox}$ ($V_{oy}$) is the in-phase (out-of-phase) component of the output signal, while $V_{Ax}$ ($V_{Ay}$) is the in-phase (out-of-phase) component of the input signal.

The measured total capacitance is related to the quantum capacitance $C_Q$ and the geometric gate ($C_g$) and parasitic ($C_p$) capacitance as~\cite{luryi1988quantum,ilani2006measurement}
\begin{equation}
    C_t = \left(\frac{1}{C_g} + \frac{1}{C_Q}\right)^{-1} + C_p,
\end{equation}
with $$C_Q = e^2 \frac{\partial n}{\partial \mu} = e^2 g(E_F)$$ $C_g$ and $C_p$ are assumed to be constant over the measurement range, so that variations in $C_t$ directly reflect changes in the electronic density of states through $C_Q$.

\begin{figure}[h]
    \centering
    \includegraphics[width=1\columnwidth]{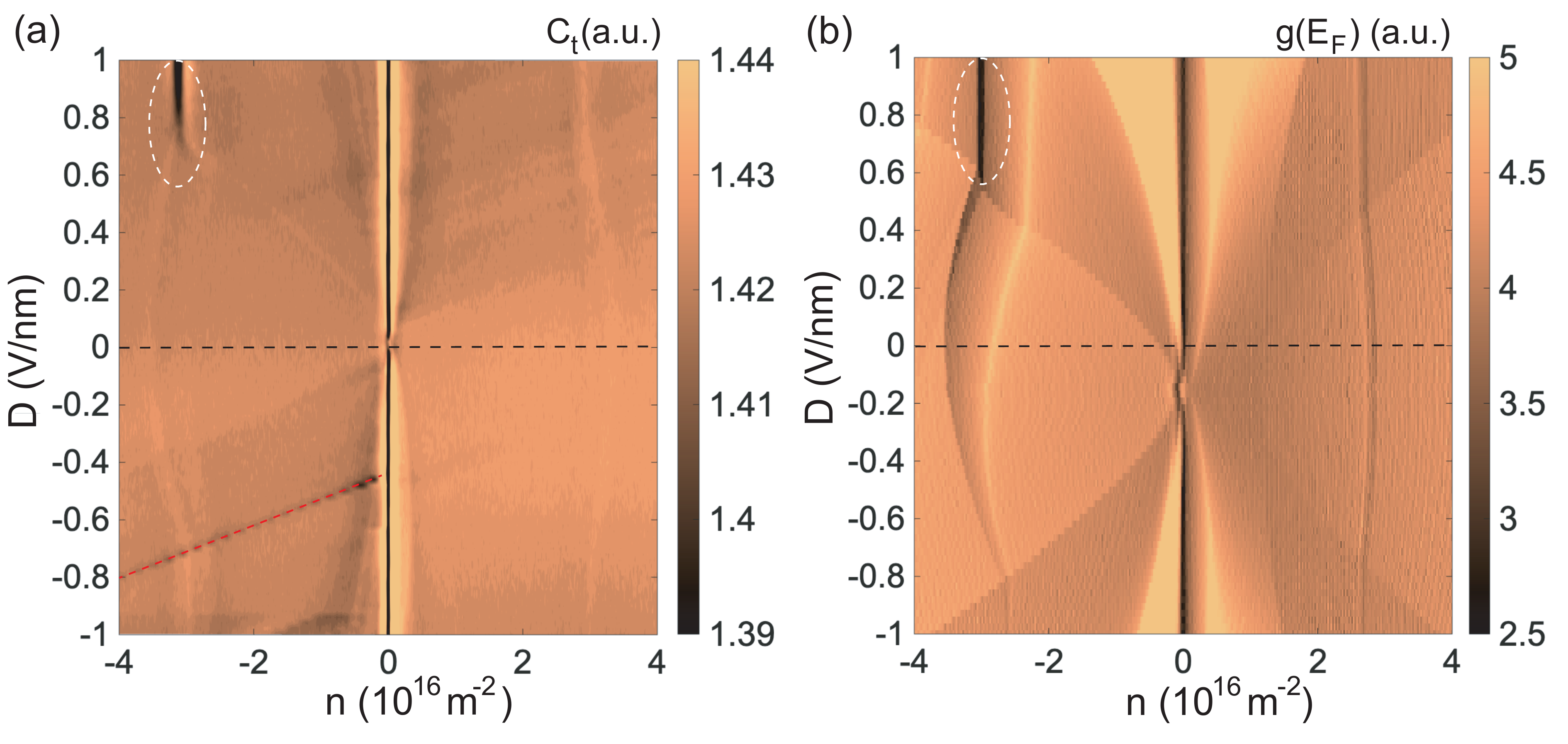}
    \caption{(a)  Measured capacitance $C_t$ as a function of $n$ and $D$. (b) Calculated density of states $g(E_F)$ as a function of $n$ and $D$.}
    \label{figS5}
\end{figure}

Fig.~\ref{figS5}(a) shows a contour plot of the measured capacitance $C_t$ as a function of $n$ and $D$. The diagonal streak observed in the data (red dotted line) arises from the formation of a p--n junction. A pronounced dip in measured capacitance $C_t$ appears at the h-SDP for $D \geq 0.7~\mathrm{V/nm}$ (white dotted ellipse). In addition, the shift of the SDP with $D$ is clearly visible. For comparison, the calculated $g(E_F)$ as a function of $n$ and $D$ is shown in Fig.~\ref{figS5}(b).

\section{Critical field $D_c$}

\begin{figure}[h]
    \centering
\includegraphics[width=\columnwidth]{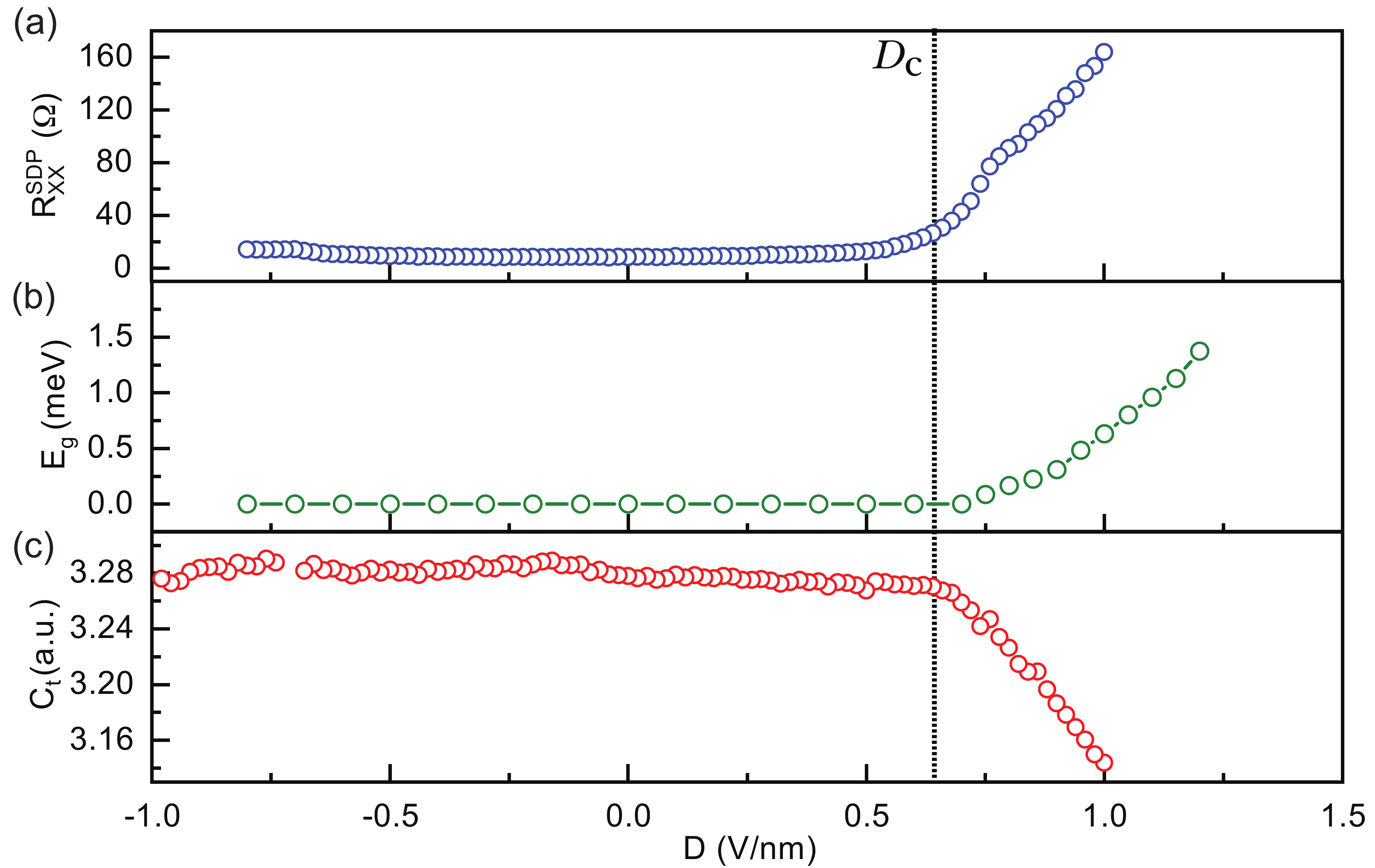}
    \caption{Plot of (a) $R^{SDP}_{xx}$, (b) $E_g$, and (c) measured capacitance $C_t$ as a function of $D$ for h-SDP. The dotted black line marks the critical value $D_c$. }
    \label{figS6}
\end{figure}

Fig.~\ref{figS6} shows plots of $R^{SDP}_{xx}$, $E_g$ and the measured capacitance $C_t$  as a function of $D$ for h-SDP. All three quantities exhibit a clear change at a common critical value, $D_c = 0.7~\mathrm{V/nm}$, indicating a shared underlying transition in the system. When $D$ is increased above $D_c$, both $R^{SDP}_{xx}$ and $E_g$ increase sharply, while $C_t$ drops suddenly. The increase in resistance and energy gap, along with the decrease in capacitance, show a change in the band structure. For $D > D_c$, the MLL band shifts to higher energies and becomes irrelevant at the SDP, hence the SDP gap arises only from the BLL band.

\section{Model Hamiltonian for ABA Trilayer Graphene on hBN}

\begin{figure}[h]
    \centering
\includegraphics[width=\columnwidth]{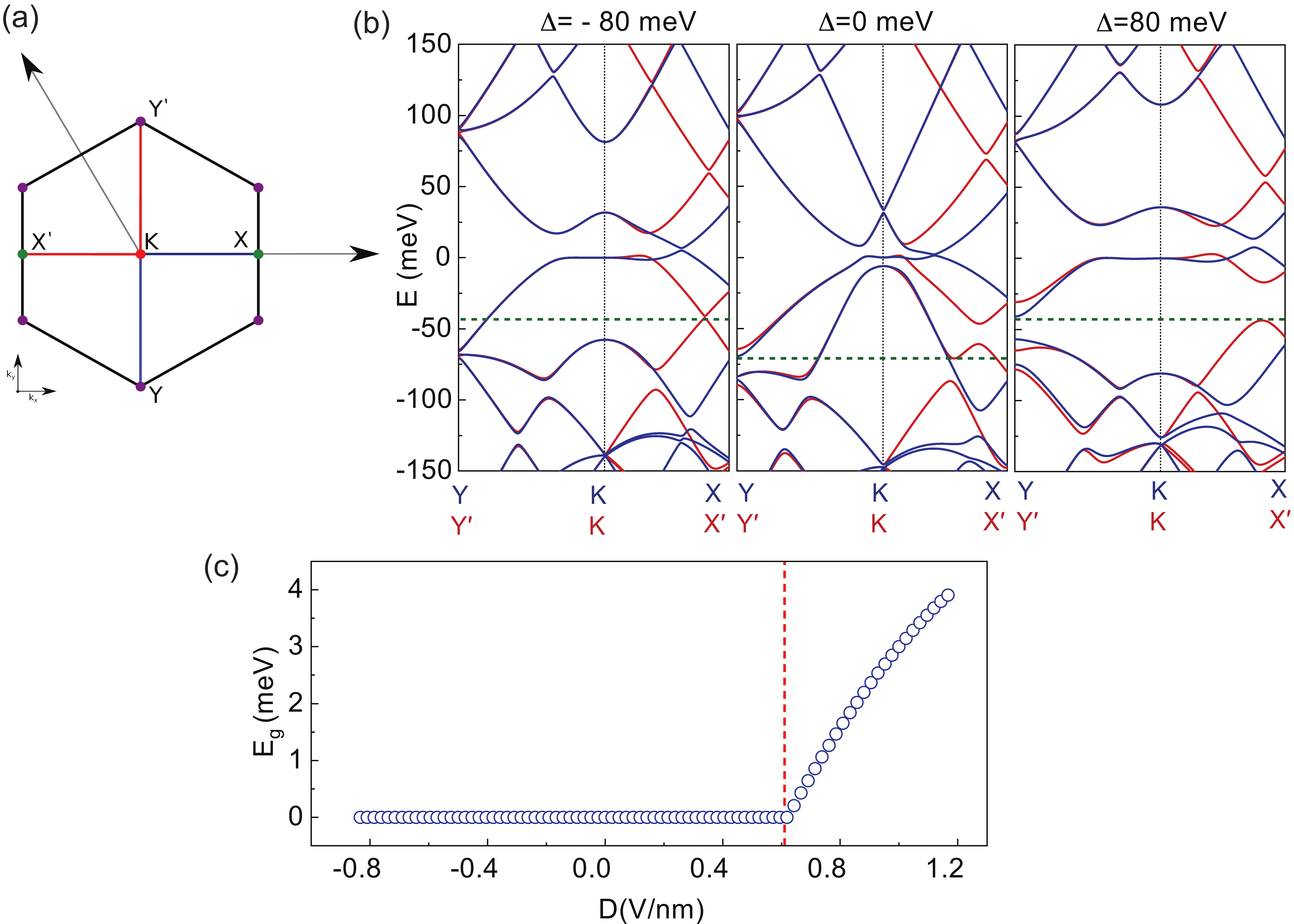}
    \caption{(a) $\mathbf{k}$-path for band structure in Brillouin zone of moir\'e lattice. (b) Calculated band structure for $\Delta=-80~\mathrm{meV}$, $\Delta=0~\mathrm{meV}$ and $\Delta=80~\mathrm{meV}$. The blue(red) curves correspond to the path $\mathrm{Y} \rightarrow \mathrm{K}\rightarrow \mathrm{X}$ ($\mathrm{Y'} \rightarrow \mathrm{K}\rightarrow \mathrm{X'}$). Green dotted line marks the position of SDP. (c) $E_g$ extracted from calculated $g(E_F)$ as a function of $D$. Red dotted line marks the position of $D_c$.}
    \label{figS7}
\end{figure}

We consider ABA-stacked trilayer graphene (TLG) in the sub-lattice basis. 
\[
(A_1,B_1,A_2,B_2,A_3,B_3),
\]
where \(A_\ell\) and \(B_\ell\) denote the two honeycomb sublattices in layer \(\ell=1,2,3\). In the continuum description \cite{PhysRevB.84.035440} near the K/K' valley , the electronic structure of pristine ABA trilayer graphene is represented by a six-band Hamiltonian composed of the standard Slonczewski--Weiss--McClure \cite{PhysRev.109.272, Dresselhaus2002} hopping processes together with an electrostatic layer-asymmetric term generated by a perpendicular displacement field \cite{PhysRevB.79.125443}. The total Hamiltonian of the isolated trilayer is written as 
\begin{equation}
H_{\mathrm{TLG}} = H_0 + H_{\Delta},
\end{equation}
where \(H_0\) contains the intralayer and interlayer hopping amplitudes and \(H_{\Delta}\) describes the potential difference between the outer layers.

On the basis specified above, the ABA trilayer graphene Hamiltonian takes the form \cite{Serbyn2013,PhysRev.104.666,Dresselhaus2002,Taychatanapat2011}
\begin{equation}
H_0 =
\begin{pmatrix}
0 & \gamma_0 t_{\mathbf{k}}^\ast & \gamma_4 t_{\mathbf{k}}^\ast & \gamma_3 t_{\mathbf{k}} & \gamma_2/2 & 0 \\
\gamma_0 t_{\mathbf{k}} & \delta & \gamma_1 & \gamma_4 t_{\mathbf{k}}^\ast & 0 & \gamma_5/2 \\
\gamma_4 t_{\mathbf{k}} & \gamma_1 & \delta & \gamma_0 t_{\mathbf{k}}^\ast & \gamma_4 t_{\mathbf{k}} & \gamma_1 \\
\gamma_3 t_{\mathbf{k}}^\ast & \gamma_4 t_{\mathbf{k}} & \gamma_0 t_{\mathbf{k}} & 0 & \gamma_3 t_{\mathbf{k}}^\ast & \gamma_4 t_{\mathbf{k}} \\
\gamma_2/2 & 0 & \gamma_4 t_{\mathbf{k}}^\ast & \gamma_3 t_{\mathbf{k}} & 0 & \gamma_0 t_{\mathbf{k}}^\ast \\
0 & \gamma_5/2 & \gamma_1 & \gamma_4 t_{\mathbf{k}}^\ast & \gamma_0 t_{\mathbf{k}} & \delta
\end{pmatrix}.
\end{equation}
Here, \(\gamma_0\) is the nearest-neighbor intralayer hopping amplitude, \(\gamma_1\) is the dominant interlayer dimer hopping, while \(\gamma_3\) and \(\gamma_4\) encode skew interlayer couplings responsible for trigonal warping and electron--hole asymmetry. The parameters \(\gamma_2\) and \(\gamma_5\) describe next-nearest-layer couplings, and \(\delta\) denotes the on-site energy shift associated with dimer sites \cite{Dresselhaus2002}. The momentum dependence is contained in the complex structure factor \(t_{\mathbf{k}}\), which in the low-energy continuum theory is taken to be $t_{\mathbf{k}} = \xi k_x + i k_y$ \cite{PhysRevB.81.115315,Ando2005,PhysRev.104.666}, with \(\mathbf{k}\) measured relative to the valley center. The valley index \(\xi=\pm1\) distinguishes the \(K\) and \(K'\) valleys and determines the chirality of the Dirac-like intralayer kinetic term.

A perpendicular electric displacement field introduces an electrostatic potential difference between the top and bottom layers. In the chosen basis, this contribution is expressed as \cite{PhysRevB.79.125443}
\begin{equation}
H_{\Delta}=\mathrm{diag}(\Delta,\Delta,0,0,-\Delta,-\Delta).
\end{equation}
This term breaks inversion symmetry across the trilayer stack and induces layer polarization, which strongly modifies the low-energy band structure. In ABA trilayer graphene, such a displacement field hybridizes the monolayer-like and bilayer-like sectors and reshapes band structure near the charge neutrality point.

When the top graphene layer is placed on or aligned with hexagonal boron nitride (hBN), the lattice mismatch and relative crystallographic orientation generate a long-wavelength moir\'e modulation. The effect of hBN on graphene can be incorporated perturbatively by integrating out the hBN degrees of freedom \cite{Moon2014, PhysRevB.81.155433, PhysRevB.87.205404, PhysRevB.89.035431, PhysRevB.86.115415}. The reduced effective Hamiltonian is written as
\begin{equation}
H^{(\mathrm{red})}_{\mathrm{G-hBN}}
= H_G + U^\dagger (-H_{\mathrm{hBN}})^{-1} U
\equiv H_G + V_{\mathrm{hBN}},
\end{equation}
where \(H_G\) is the graphene Hamiltonian, \(H_{\mathrm{hBN}}\) is the substrate Hamiltonian, and \(U\) denotes the graphene--hBN interlayer coupling matrix. The second term acts as an effective periodic potential \(V_{\mathrm{hBN}}\) induced on graphene. This form corresponds to a Schrieffer--Wolff-type reduction \cite{Moon2014,PhysRevB.87.205404} in which the hBN bands, assumed to lie far from the graphene low-energy window, are integrated out to produce a static self-energy correction for the graphene layer.

The moir\'e potential is represented as a sum of a uniform term and the leading Fourier harmonics associated with the moir\'e reciprocal lattice vectors. Its explicit form is \cite{Moon2014}
\begin{align}
V_{\mathrm{hBN}}(\mathbf{r})
&=
V_0
\begin{pmatrix}
1 & 0\\
0 & 1
\end{pmatrix}
\nonumber\\
&\quad+
\Bigg\{
V_1 e^{i\xi\psi}
\Bigg[
\begin{pmatrix}
1 & \omega^{-\xi}\\
1 & \omega^{-\xi}
\end{pmatrix}
e^{i\xi \mathbf{G}_1^M\cdot \mathbf{r}}
+
\begin{pmatrix}
1 & \omega^{\xi}\\
\omega^{\xi} & \omega^{-\xi}
\end{pmatrix}
e^{i\xi \mathbf{G}_2^M\cdot \mathbf{r}}
\nonumber\\
&\qquad\qquad\qquad\qquad+
\begin{pmatrix}
1 & 1\\
\omega^{-\xi} & \omega^{-\xi}
\end{pmatrix}
e^{-i\xi(\mathbf{G}_1^M+\mathbf{G}_2^M)\cdot \mathbf{r}}
\Bigg]
+\mathrm{H.c.}
\Bigg\},
\end{align}
where \(V_0\) denotes the spatially uniform part of the substrate-induced potential and \(V_1\) sets the amplitude of the leading moir\'e modulation. The phase \(\psi\) parametrizes the relative complex phase of the first-shell harmonics, \(\omega=e^{2\pi i/3}\) is the cubic root of unity associated with the threefold rotational structure of the honeycomb lattice, and \(\mathbf{G}_1^M\), \(\mathbf{G}_2^M\) are primitive reciprocal vectors of the moir\'e superlattice.
The parameters of the moir\'e potential are obtained from an electric-field-dependent model, where the applied perpendicular displacement field is denoted by \(\Delta\). In the numerical implementation, the substrate-induced onsite potentials for nitrogen and boron sites are taken to depend linearly on the displacement field as
\begin{equation}
V_N^{\text{eff}} = V_N + \Delta, 
\qquad
V_B^{\text{eff}} = V_B + \Delta,
\end{equation}
where $V_N=-1.4$ eV and $V_B=3.34$ eV \cite{PhysRevB.81.155433}. The intrinsic graphene-hBN coupling strength is taken to be $u_0 = 0.127 \text{eV}$.

The uniform and modulated components of the moir\'e potential are defined as
\begin{equation}
V_0 = -3 u_0^2 \left( \frac{1}{V_N^{\text{eff}}} + \frac{1}{V_B^{\text{eff}}} \right),
\end{equation}
and
\begin{equation}
V_{1}e^{i\xi\psi} = - u_0^2 \left( \frac{1}{V_N^{\text{eff}}} + \frac{e^{i \xi 2\pi/3}}{V_B^{\text{eff}}} \right).
\end{equation}

This construction provides a convenient way to incorporate the dependence of the moir\'e potential on the external electric field within the continuum model.
For the trilayer graphene/hBN heterostructure, the substrate couples directly only to the graphene layer adjacent to hBN. The complete continuum Hamiltonian is therefore written in layer-block form as
\begin{equation}
H=
\begin{pmatrix}
H_{\mathrm{gr}_1}+V_{\mathrm{hBN}} & H_{12} & H_{13}\\
H_{12}^{\dagger} & H_{\mathrm{gr}_2} & H_{23}\\
H_{13}^{\dagger} & H_{32}^{\dagger} & H_{\mathrm{gr}_3}
\end{pmatrix}.
\end{equation}
Here \(H_{\mathrm{gr}_\ell}\) denotes the intralayer Dirac Hamiltonian for graphene layer \(\ell\), while \(H_{12}\), \(H_{23}\), and \(H_{13}\) represent the interlayer tunneling matrices appropriate to ABA stacking. In this representation, the moir\'e potential enters only at layer \(\ell=1\) diagonal block.

This model captures the combined effects of ABA interlayer hybridization, remote hopping processes, electrostatic asymmetry, and the hBN-induced moir\'e superlattice potential within a single continuum framework. Physically, the hBN substrate reconstructs the spectrum of the proximate layer and, through interlayer coupling, influences the entire trilayer electronic structure. As a result, the low-energy bands form moir\'e minibands.
The tight-binding parameters entering the ABA trilayer graphene Hamiltonian are taken from the standard Slonczewski--Weiss--McClure parametrization. In this work, we use the values listed in Table~\ref{tab:parameters}.

\begin{table*}[t]
\centering
\begin{tabular}{ccccccc}
\hline
$\gamma_0$ & $\gamma_1$ & $\gamma_2$ & $\gamma_3$ & $\gamma_4$ & $\gamma_5$ & $\delta$ \\
\hline
3.1 & 0.39 & -0.028 & 0.315 & -0.041 & 0.05 & 0.046 \\
\hline
\end{tabular}
\caption{Slonczewski--Weiss--McClure parameters (in eV) used for ABA trilayer graphene. \cite{Taychatanapat2011}\label{tab:parameters}}
\end{table*}

We subsequently calculated the band structure along two different paths, as shown in Fig.~\ref{figS7}(a), for three values of $\Delta = -80, 0, 80$ meV, corresponding to negative, zero, and positive values of $D$, respectively. For $D > 0$ (Fig.~\ref{figS7}(b)), a gap opens on the hole side at the Y/Y$'$ points and at intermediate points along the K to X$'$ path. In contrast, for electron doping, the magnitude of the gap is significantly smaller, similar to the graphene/hBN moir\'e case \cite{Moon2014,PhysRevB.86.115415}. 

For $D < 0$, the gap opening at Y/Y$'$ (and close to X/X$'$)  is much smaller, and consequently no gap is observed in the density of states (see Fig.~\ref{figS5}(b). This is due to the smaller projection of the moir\'e layer onto the states near the secondary Dirac point (SDP), as explained in the next section. 

We further calculated the variation of the gap, defined as $E_g = \min\{E_{v\{\mathbf{k}\}}\}_{v=1} - \max\{E_{v\{\mathbf{k}\}}\}_{v=2}$, as a function of $\Delta$. The trend, observed in Fig.~\ref{figS7}(c), is in excellent agreement with the trend observed in the experiment (see Fig.~3 in the main text).

\section{Effect of trigonal warping and Layer polarization in the presence of moir\'e potential}

\begin{figure}[h]
    \centering
\includegraphics[width=1.0\columnwidth]{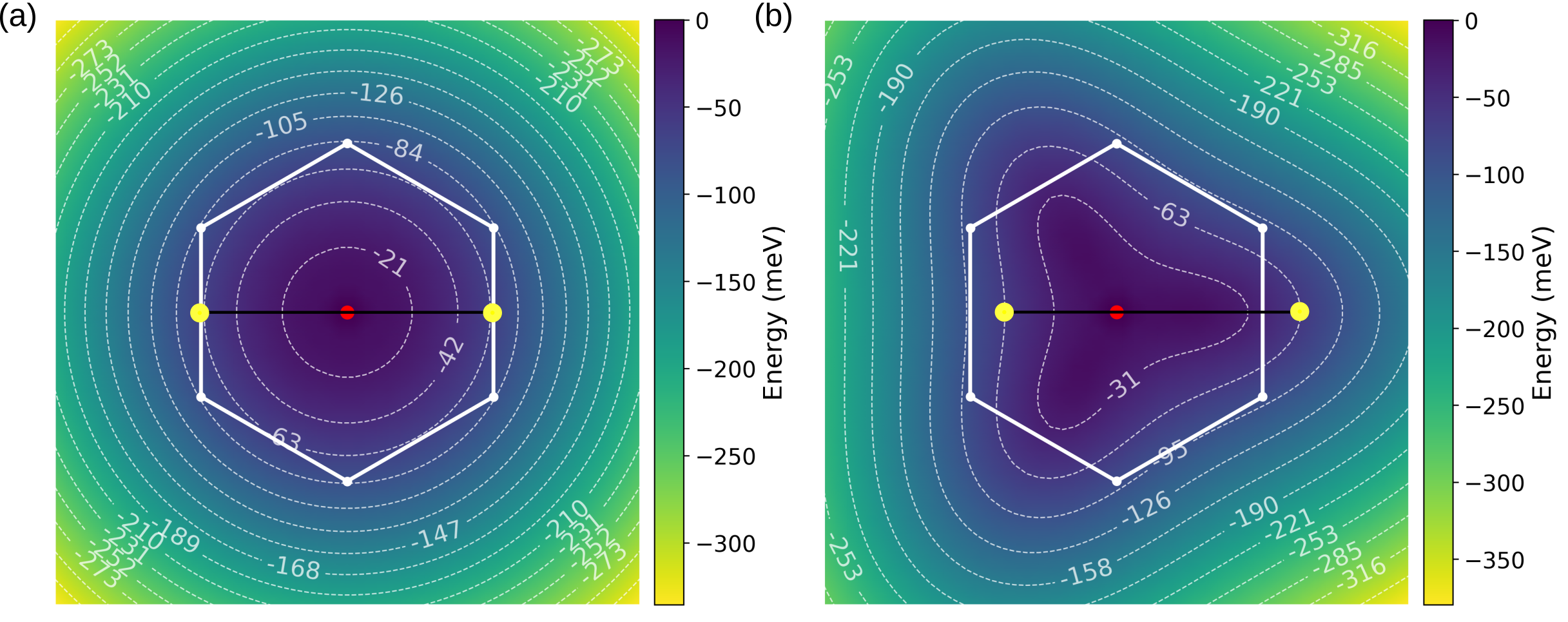}
\caption{ Iso-energy contour plots for band structure within K valley: (a) in the absence of trigonal warping, and (b) in the presence of trigonal warping. The yellow dot indicates the location within the Brillouin zone where a band gap emerges as a result of the moir\'e potential. The hexagon represents the moir\'e Brillouin zone corresponding to a twisted structure with a twist angle of $0.55^\circ$.
}
\label{Fig_trigonal_warping}
\end{figure}
Trigonal warping in multilayer graphene exerts a pronounced influence on the low-energy electronic band structure \cite{PhysRevB.80.165409}. To elucidate the origin of the small band gap observed at the secondary Dirac point, we examine the role of trigonal warping in the presence of a moir\'e potential. The inclusion of trigonal warping reduces the symmetry of the band structure within an individual valley from $C_6$ to $C_3$, giving rise to characteristic triangular iso-energy contours in reciprocal space.

In Fig.~\ref{Fig_trigonal_warping}(a), we plot the computed iso-energy contours of trilayer graphene in the absence of a moir\'e potential, with $\gamma_3 = 0$ (i.e., neglecting trigonal warping). When the moir\'e potential is introduced, gaps are expected to open at points on the iso-energy contours that are connected by a moir\'e reciprocal lattice vector $\{\mathbf{G}^M\}$. In this case, the energies at the Y(Y$'$) points are equal, as are those at the X(X$'$) points. Since these points are connected by $\{\mathbf{G}^M\}$, gap openings are expected at these high-symmetry locations, as indicated by the yellow dots for X(X$'$) in Fig.~\ref{Fig_trigonal_warping}(a). The energy difference between the Y(Y$'$) and X(X$'$) points is approximately 20~meV.

In contrast, when trigonal warping is included by setting $\gamma_3 = 0.315$~eV, the energies at X and X$'$ are no longer degenerate (see Fig.~\ref{Fig_trigonal_warping}(b)). Consequently, the points on the iso-energy contours connected by the moir\'e vector $\{\mathbf{G}^M\}$ shift away from the high-symmetry X(X$'$) points, and the corresponding gap openings occur slightly displaced from these locations. In this case, the energy difference between the Y(Y$'$) and X(X$'$) points increases to approximately 30~meV. Therefore, trigonal warping not only shifts the gap opening away from high-symmetry points but also enhances the energy separation between the secondary Dirac points (SDPs), leading to a smaller effective gap observed in transport measurements.

\begin{figure}[t]
    \centering
    \includegraphics[width=1.0\linewidth]{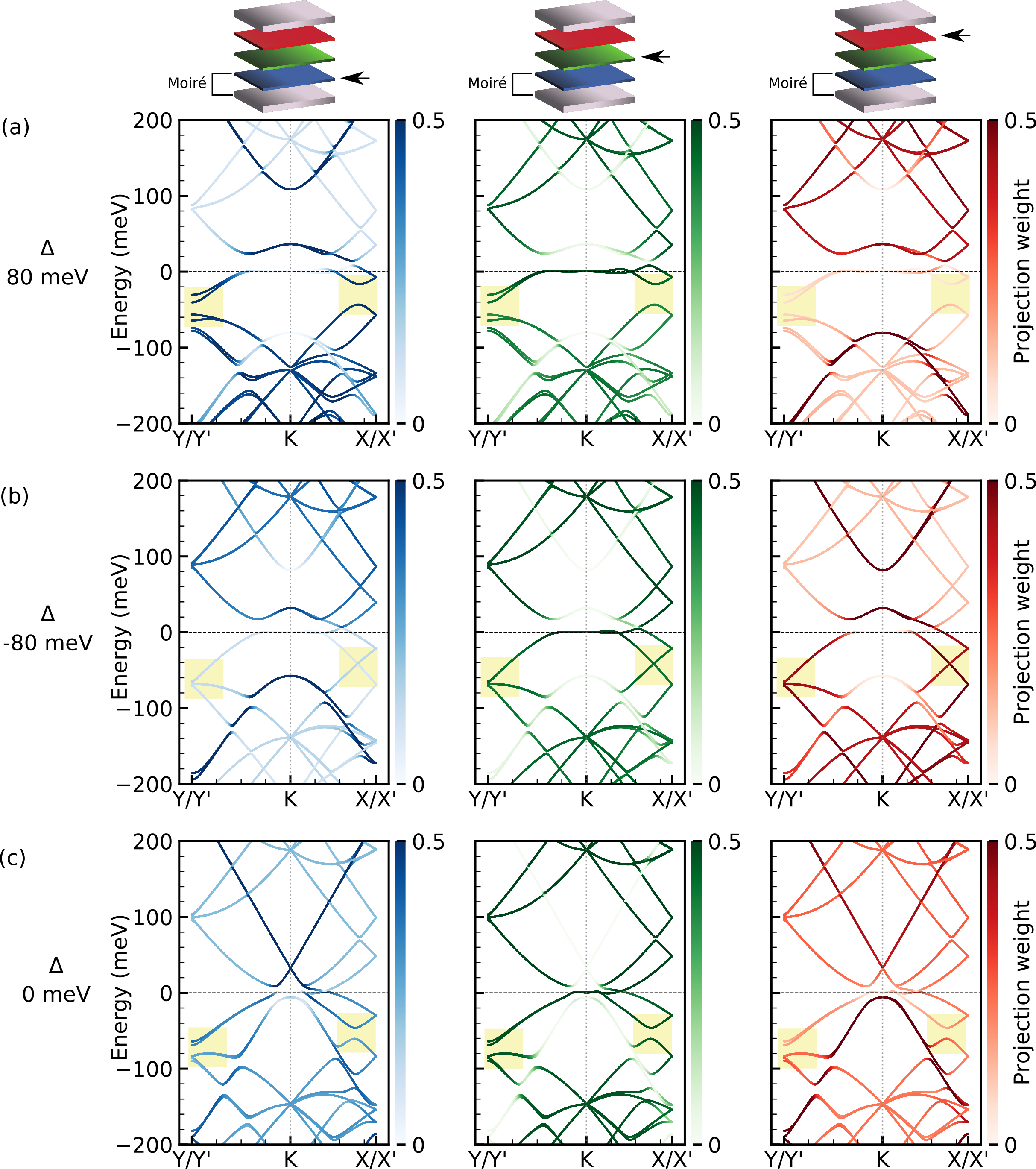}
    \caption{Layer-projected band structures for (a) $\Delta = 80$ meV, (b) $\Delta = -80$ meV, and (c) $\Delta = 0$ meV. The yellow shaded regions highlight the secondary Dirac points (SDPs) near the Y/Y$'$ and X/X$'$ points. The color scale represents the layer projection weight, defined for each layer $l$ as $\langle \psi_{n\mathbf{k}} \vert P_l \vert \psi_{n\mathbf{k}} \rangle$, where $P_l$ is the projector onto layer $l$. In panels (a)–(c), blue denotes the layer coupled to the hBN moiré potential, green corresponds to the middle layer, and red represents the top layer.}
    
    \label{fig:Projection}
\end{figure}

The band gap is also strongly influenced by layer polarization. To elucidate the effect of the external field on both the layer polarization and the gap, we compute the layer-projected band structure, shown in Fig.~\ref{fig:Projection}.
In Fig.~\ref{fig:Projection}(a), corresponding to $\Delta = 80$ meV, the bottom (moir\'e-coupled) layer and the middle layer carry the dominant projection weight near the secondary Dirac points (SDPs), highlighted in yellow, while the top (non-moir\'e) layer contributes negligibly. In this case, a clear gap opening is observed in the highlighted regions.

By contrast, in Fig.~\ref{fig:Projection}(b) for $\Delta = -80$ meV, the bottom layer exhibits very small projection weight near the SDPs, which leads to a significantly reduced gap.
Finally, in Fig.~\ref{fig:Projection}(c) for $\Delta = 0$, all three layers contribute substantially near the SDPs at both X(X$'$) and Y(Y$'$). Although gaps are present at these points, the system remains metallic due to the presence of single-layer-like bands.
Together with trigonal warping, these factors explain the reduced gap and lower resistivity observed in trilayer graphene moir\'e systems compared to both single-layer graphene and bilayer graphene/hBN moir\'e systems.
Furthermore, as trigonal warping lowers the symmetry of the band structure, the twist-angle dependence of the gap becomes nontrivial, and the observed gap can exhibit a strong sensitivity to the relative orientation of the moir\'e potential with respect to trilayer graphene.

\section{Details of calculations}
We first determine a commensurate structure near a twist angle of approximately $0.55^\circ$. For this twist angle, we obtain $(c_1, c_2) = (28, 28)$ for the moir\'e lattice vector $\mathbf{L}_1 = c_1 \mathbf{a}_1 + c_2 \mathbf{a}_2$, where $\mathbf{a}_i$ are the lattice vectors of trilayer graphene. This choice corresponds to a commensurate structure with minimal strain and yields a moir\'e length of approximately $118.8$ \AA. We only did calculations for K valley as K' valley is related to K valley by time reversal symmetry. 

In the carrier density calculations, we rescale the hole doping density $n$ such that the theoretical value $n_s = 3.28 \times 10^{16} \mathrm{m}^{-2}$ matches the experimentally observed value $n_s = 3.05 \times 10^{16} \mathrm{m}^{-2}$. This procedure is justified because the scaling is applied uniformly across all hole densities. We find that this rescaling leads to excellent agreement between theoretical predictions and experimental results, further supporting its validity.

For the density of states calculations, the entire Brillouin zone is sampled using a $100 \times 100$ $k$-point grid, which ensures convergence of both the density of states and the band gap.

\clearpage

\bibliography{arXiv}

\end{document}